\documentclass[fleqn,usenatbib,usedcolumn]{mnras}

\usepackage[british]{babel}

\usepackage{ae,aecompl}
\usepackage[T1]{fontenc}
\usepackage{booktabs}

\usepackage{graphicx}
\usepackage{amsmath}
\usepackage{amssymb}

\newcommand{\bagpipes}{\textsc{Bagpipes}}
\newcommand*\samethanks[1][\value{footnote}]{\footnotemark[#1]}

\title[Quiescent galaxy SFHs at $z\gtrsim1$]{The VANDELS survey: the star-formation histories of massive quiescent galaxies at 1.0~<~z~<~1.3}

\author[A. C. Carnall et al.]{A. C. Carnall$^{1}$\thanks{E-mail: adamc@roe.ac.uk},
R. J. McLure$^{1}$,
J. S. Dunlop$^{1}$,
F. Cullen$^{1}$,
D. J. McLeod$^{1}$,
\newauthor V. Wild$^{2}$,
B. D. Johnson$^{3}$,
S. Appleby$^{1}$,
R. Dav\'e$^{1, 4, 5}$,
R. Amorin$^{6, 7}$,
M. Bolzonella$^{8}$,
\newauthor M. Castellano$^{9}$,
A. Cimatti$^{10, 11}$,
O. Cucciati$^{8}$,
A. Gargiulo$^{12}$,
B. Garilli$^{12}$,
\newauthor F. Marchi$^{9}$,
L. Pentericci$^{9}$,
L. Pozzetti$^{8}$,
C. Schreiber$^{13}$,
M. Talia$^{8,10}$,
G. Zamorani$^{8}$
\\
\\
$^{1}$ SUPA\thanks{Scottish Universities Physics Alliance}, Institute for Astronomy, University of Edinburgh, Royal Observatory, Edinburgh EH9 3HJ, UK \\
$^{2}$ SUPA\samethanks, School of Physics and Astronomy, University of St. Andrews, North Haugh, St. Andrews KY16 9SS, UK \\
$^{3}$ Harvard-Smithsonian Center for Astrophysics, 60 Garden St., Cambridge, MA 02138, USA\\
$^{4}$ University of the Western Cape, Bellville, Cape Town 7535, South Africa \\
$^{5}$ South African Astronomical Observatories, Observatory, Cape Town 7925, South Africa \\
$^{6}$ Instituto de Investigaci\'on Multidisciplinar en Ciencia y Tecnolog\'ia, Universidad de La Serena, Ra\'ul Bitr\'an 1305, La Serena, Chile \\
$^{7}$ Departamento de F\'isica y Astronom\'ia, Universidad de La Serena, Av. Juan Cisternas 1200 Norte, La Serena, Chile \\
$^{8}$ INAF - OAS Bologna, Via P. Gobetti 93/3, I-40129, Bologna, Italy \\
$^{9}$ INAF - Osservatorio Astronomico di Roma, Via Frascati 33, I-00078 Monteporzio Catone, Italy \\
$^{10}$ University of Bologna, Department of Physics and Astronomy (DIFA), Via Gobetti 93/2, I-40129, Bologna, Italy \\
$^{11}$ INAF - Osservatorio Astrofisico di Arcetri, Largo E. Fermi 5, I-50125, Firenze, Italy \\
$^{12}$ INAF - IASF Milano, Via A. Corti 12, I-20133, Milano, Italy \\
$^{13}$ Department of Physics, University of Oxford, Keble Road, Oxford OX1 3RH, UK
}
\date{Accepted XXX. Received YYY; in original form ZZZ}
\pubyear{2019}

\begin{document}
\label{firstpage}
\pagerange{\pageref{firstpage}--\pageref{lastpage}}
\maketitle

\begin{abstract}
We present a Bayesian full-spectral-fitting analysis of 75 massive ($M_* > 10^{10.3} \mathrm{M_\odot}$) UVJ-selected galaxies at redshifts of $1.0 < z < 1.3$, combining extremely deep rest-frame ultraviolet spectroscopy from VANDELS with multi-wavelength photometry. By the use of a sophisticated physical plus systematic uncertainties model, constructed within the \bagpipes\ code, we place strong constraints on the star-formation histories (SFHs) of individual objects. We firstly constrain the stellar mass vs stellar age relationship, finding a steep trend towards earlier average formation time with increasing stellar mass (downsizing) of $1.48^{+0.34}_{-0.39}$ Gyr per decade in mass, although this shows signs of flattening at $M_* > 10^{11} \mathrm{M_\odot}$. We show that this is consistent with other spectroscopic studies from $0 < z < 2$. This relationship places strong constraints on the AGN-feedback models used in cosmological simulations. We demonstrate that, although the relationships predicted by \textsc{Simba} and \textsc{IllustrisTNG} agree well with observations at $z=0.1$, they are too shallow at $z=1$, predicting an evolution of $\lesssim0.5$ Gyr per decade in mass. Secondly, we consider the connections between green-valley, post-starburst and quiescent galaxies, using our inferred SFH shapes and the distributions of galaxy physical properties on the UVJ diagram. The majority of our lowest-mass galaxies ($M_* \sim 10^{10.5} \mathrm{M_\odot}$) are consistent with formation in recent ($z<2$), intense starburst events, with timescales of $\lesssim500$ Myr. A second class of objects experience extended star-formation epochs before rapidly quenching, passing through both green-valley and post-starburst phases. The most massive galaxies in our sample are extreme systems: already old by $z=1$, they formed at $z\sim5$ and quenched by $z=3$. However, we find evidence for their continued evolution through both AGN and rejuvenated star-formation activity.
\end{abstract}

\section{Introduction}\label{sect:intro}

Understanding the origin of the colour bimodality in the local galaxy population remains one of the most important goals of extragalactic astronomy. The emerging picture is one in which feedback of energy from baryonic processes plays a central role in the quenching of star-formation activity, with supernovae and active galactic nuclei (AGN) thought to dominate at lower and higher masses respectively. A range of other factors are also thought to play important roles, for example mergers and environmental effects such as ram-pressure stripping. However, we still lack a detailed physical description of these and other relevant processes.

The inherent complexity of galaxy formation physics means that large-scale numerical simulations are required to connect theoretical models with observable properties. The extreme computational expense of such simulations precludes, for the present, a Bayesian statistical approach to parameter estimation and model selection in this context. Instead, predictions from individual simulations are made for a number of observable properties, such as the distributions of galaxy colours, stellar masses and star-formation rates (SFRs), which can be compared with observational results. These comparisons provide qualitative insights as to how the underlying physical model should be refined (e.g. \citealt{Dave2017}; \citealt{Trayford2017}; \citealt{Nelson2018}; \citealt{Donnari2018}; \citealt{Cochrane2018}; \citealt{Baes2019}). 

However, numerical simulations have now reached a level of complexity such that a range of models with varying physical ingredients can produce good approximations of the standard set of well-constrained observable properties. The challenge for new observational studies therefore is to provide precise measurements of a wider range of physical parameters which are highly constraining on galaxy formation models, such as stellar ages, star-formation histories (SFHs), stellar and nebular metallicities and levels of dust attenuation. Crucially, these studies must be performed on large and representative samples of galaxies. They must also be extended to high redshift, in order to constrain the evolution of galaxy properties across cosmic time.

An area of particular theoretical interest in recent years has been the AGN-feedback processes which quench star-formation in the most massive galaxies (e.g. \citealt{Dave2016, Dave2019}; \citealt{Weinberger2017}; \citealt{Pillepich2018}; \citealt{Nelson2019}). These processes should leave strong imprints, not just on the bright end of the galaxy luminosity function, but also on the star-formation histories of massive quiescent galaxies (e.g. \citealt{Croton2006}). Additional constraints are also available from the physical properties of galaxies transitioning between the star-forming and quiescent populations. If strong constraints can be placed on these more subtle indicators, it will be possible to begin ruling out models which are capable of matching simpler observables.

A huge literature exists on the stellar ages of massive quiescent galaxies (e.g. \citealt{Heavens2000, Heavens2004}; \citealt{Panter2003, Panter2007};  \citealt{Cimatti2004, Cimatti2008}; \citealt{Gallazzi2005, Gallazzi2014}; \citealt{Daddi2005}; \citealt{Onodera2012, Onodera2015}; \citealt{Jorgensen2013}; \citealt{Whitaker2013}; \citealt{Choi2014, Choi2019}; \citealt{Fumagalli2016}; \citealt{Pacifici2016}; \citealt{Citro2016}; \citealt{Siudek2017}; \citealt{Carnall2018}; \citealt{Belli2018}; \citealt{Estrada-Carpenter2019}). However, these measurements are challenging to make for several reasons.

As stellar populations age their luminosities fall rapidly, meaning that the evidence of earlier star-formation episodes can easily be lost in the glare of younger stellar populations. Additionally, galaxy spectral energy distributions (SEDs) suffer from strong, interrelated degeneracies between different physical properties, such as the age-metallicity-dust degeneracy (e.g. \citealt{Papovich2001}; \citealt{Lee2007}; \citealt{Conroy2013}). Because of these issues, photometric data often fail to strongly constrain galaxy physical parameters, meaning the applied priors can significantly impact the results obtained (e.g. \citealt{carnall2019}; \citealt{Leja2019a}). Finally, significant systematic uncertainties exist in the empirical models used to interpret observational data (e.g. \citealt{Han2019}).

The situation has been improved in the local Universe by the advent of large systematic surveys designed to obtain high signal-to-noise ratio (SNR) continuum spectroscopy. These data are more strongly constraining on subtle galaxy physical parameters (e.g. \citealt{Pacifici2012}; \citealt{Thomas2017}), however, until recently such data has been extremely scarce at higher redshifts. Additionally, interpreting spectroscopic data in a way that makes full use of the available information is challenging, both in terms of the complexity of the required models and the computational expense of fitting these models to data (see Section \ref{subsect:fitting_history}). The use of simplifying assumptions to reduce the complexity of the problem, for example by fixing nuisance parameters to fiducial values, typically leads to biases in derived physical parameter values and to underestimated uncertainties (e.g. \citealt{Pacifici2015}; \citealt{Iyer2017}).

Despite these challenges, a consensus has emerged around several important results. Firstly, at fixed observed redshift, less-massive galaxies are found to have younger stellar populations than their more massive counterparts (e.g. \citealt{Gallazzi2005, Gallazzi2014}; \citealt{Pacifici2016}; \citealt{Carnall2018}). This is often referred to as downsizing, or mass-accelerated evolution. Secondly, at fixed stellar mass, a trend towards lower average formation redshift is found with decreasing observed redshift. A combination of factors contribute to this, including new galaxies joining the red sequence (e.g. \citealt{Brammer2011}; \citealt{Muzzin2013}; \citealt{Tomczak2014}), mergers (e.g. \citealt{Khochfar2009, Khochfar2011}; \citealt{Emsellem2011}), and periods of rejuvenated star-formation activity (e.g. \citealt{Belli2017}). 

Finally, there is considerable evidence for at least two distinct quenching mechanisms with different timescales, which change in relative importance with observed redshift (e.g. \citealt{Schawinski2014}; \citealt{Schreiber2016}; \citealt{Wild2016}; \citealt{Maltby2018}; \citealt{Carnall2018}; \citealt{Belli2018}). Rapid quenching, often associated with post-starburst galaxies (e.g. \citealt{Wild2009}), is thought to dominate at high redshift ($z\gtrsim1$), whereas slower quenching, associated with green-valley galaxies, is thought to dominate at lower redshift ($z\lesssim1$). Whilst these three fundamental results have gained broad acceptance, precise quantitative measurements are still lacking, in particular at high redshift.

Within the last year, two new, large, high-redshift spectroscopic surveys have been completed: Lega-C (\citealt{vanderwel2016}; \citealt{Straatman2018}) and VANDELS (\citealt{McLure2018b}; \citealt{Pentericci2018}). These surveys have greatly expanded the availability of high-SNR continuum spectroscopy within the first eight billion years of cosmic history, providing new opportunities for placing strong constraints on subtle galaxy physical parameters (e.g. \citealt{Cullen2017, Cullen2019}; \citealt{Wu2018a, Wu2018b}; \citealt{Chauke2018}).

In parallel, a new generation of spectral modelling and fitting tools has been developed (e.g. \citealt{Chevallard2016}; \citealt{Leja2017}; \citealt{Carnall2018}; Johnson et al. in prep). These codes include complex, flexible physical models of the kind necessary to reproduce the properties of observed spectroscopic data, and make use of modern computational and statistical methods to fit these models to data within a fully Bayesian framework. This allows the recovery of full posterior distributions for physical parameters, meaning realistic uncertainties can be obtained, including an understanding of complex, multi-parameter degeneracies.

In this work we present the first analysis of a sample of extremely deep rest-frame near-ultraviolet spectra from VANDELS. Our targets are 75 UVJ-selected galaxies with stellar masses of log$_{10}(M_*$/M$_\odot$) > 10.3 at redshifts of $1.0 < z < 1.3$. We analyse our spectra, in parallel with multi-wavelength photometry, using Bayesian Analysis of Galaxies for Physical Inference and Parameter EStimation (\bagpipes; \citealt{Carnall2018}). The \bagpipes\ code is used to fit a complex physical plus systematic uncertainties model, allowing us to obtain strong yet realistic constraints on the physical parameters of our target galaxies (see Section \ref{sect:fitting}). In particular we will discuss their SFHs, quantifying the downsizing trend at $z\sim1$ and comparing our results to predictions from cosmological simulations. We will also consider the properties of the post-starburst and green-valley galaxies in our sample, in an attempt to understand the evolutionary pathways of galaxies towards the red sequence.

The structure of this work is as follows. In Section \ref{sect:data}, we introduce VANDELS, and give details of the selection of our sample. Then, in Section \ref{sect:model}, we give details of the physical model we construct within \bagpipes\ to describe our targets. In Section \ref{sect:fitting} we discuss spectral fitting approaches, and introduce our spectroscopic plus photometric fitting methodology. We present our results in Section \ref{sect:results}, discuss these results in Section \ref{sect:discussion}, and present our conclusions in Section \ref{sect:conclusion}. All magnitudes are quoted in the AB system. For cosmological calculations we adopt $\Omega_M = 0.3$, $\Omega_\Lambda = 0.7$ and $H_0$ = 70 $\mathrm{km\ s^{-1}\ Mpc^{-1}}$. All times, $t$, are measured forwards from the beginning of the Universe. For posterior distributions we quote 50\textsuperscript{th} percentile values and 16\textsuperscript{th}$-$84\textsuperscript{th} percentile ranges. We assume a \cite{Kroupa2001} initial mass function.

\begin{table*}
  \caption{Details of the sample of 75 VANDELS objects considered in this work, selected as described in Section \ref{sect:data}. Spectroscopic redshifts were measured as described in \protect \cite{Pentericci2018}. Effective radii are taken from \protect\cite{vanderwel2014}. Stellar masses and formation redshifts are derived from the analysis described in Section \ref{subsect:fitting_final}. The full table is available as supplementary online material.}
\begingroup
\setlength{\tabcolsep}{6pt} 
\renewcommand{\arraystretch}{1.2} 
\begin{tabular}{lccccccccc}
\hline
Object ID  & RA / deg & DEC\,/\,deg & $i$ & $H$ & $z_\mathrm{spec}$ & $r_e$\,/\,arcsec & SNR\textsubscript{7500\AA}\,/\,\AA$^{-1}$ & log$_{10}(M_*$/M$_\odot$) & $z_\mathrm{form}$\\
\hline
UDS-HST-003416&34.5842&$-5.2585$&24.03&21.47&1.261&0.54&6.5&$11.13^{+0.11}_{-0.12}$&$3.2^{+0.70}_{-0.70}$ \\
UDS-HST-004029&34.5627&$-5.2557$&22.43&20.11&1.110&0.57&18.3&$11.49^{+0.15}_{-0.12}$&$2.8^{+0.60}_{-0.20}$ \\
UDS-HST-004674&34.5525&$-5.2503$&23.60&21.75&1.290&0.21&15.0&$10.79^{+0.09}_{-0.07}$&$2.7^{+0.30}_{-0.20}$ \\
UDS-HST-006039&34.5254&$-5.2427$&23.34&21.03&1.147&0.34&8.2&$11.24^{+0.18}_{-0.13}$&$3.9^{+0.90}_{-1.20}$ \\
UDS-HST-007598&34.5152&$-5.2345$&24.23&21.91&1.170&-&8.6&$10.89^{+0.16}_{-0.12}$&$2.5^{+0.20}_{-0.30}$ \\

& & & & & \textbf{\ldots} & & & \\
\hline
\end{tabular}
\endgroup
\label{table:objects}
\end{table*}

\section{VANDELS Data and Sample Selection}\label{sect:data}

VANDELS \citep{McLure2018b, Pentericci2018} is a large, recently completed ESO Public Spectroscopic Survey using the VIMOS instrument on the VLT at Paranal Observatory. The survey targeted 2106 high-redshift galaxies in the UKIDSS Ultra-Deep Survey (UDS) field and Chandra Deep Field South (CDFS). Whilst 87 per cent of the VANDELS targets are star-forming galaxies at $z>2.4$, the final 13 per cent are massive, UVJ-selected passive galaxies at $1.0 < z < 2.5$. In this section we describe our data, as well as the selection of our mass complete sample of 75 objects from the VANDELS DR2 public release\footnote{\url{https://www.eso.org/sci/publications/announcements/sciann17139.html}}

\subsection{Photometric catalogues and parent sample}\label{subsect:data_photometry}

The VANDELS photometric catalogues and sample selection procedure are both described in full  in \cite{McLure2018b}. Here we present a brief summary of the key points relevant to this work. Both sets of VANDELS pointings in UDS and  CDFS are centred on the CANDELS fields \citep{Grogin2011, Koekemoer2011}. Because the VIMOS field of view is larger than the areas imaged by CANDELS, the VANDELS photometric catalogues were supplemented with a variety of ground-based public imaging data. Because of this, the VANDELS sample is drawn from four different photometric catalogues, each spanning a UV-NIR observed wavelength range from $\sim0.3{-}5$ $\mu$m.

Each of these catalogues was subjected to an extensive SED fitting campaign to construct derived-parameter catalogues, including robust photometric redshifts, $z_\mathrm{phot}$, stellar masses and rest-frame magnitudes. The initial VANDELS sample selection was performed using a 2015 version of these catalogues, which has since been supplemented by deeper data. The photometric data used in this work, as well as the stellar masses used in Section \ref{subsect:data_sample}, come from the v1.0 internal catalogues, the final versions of which will be made public as part of the final VANDELS data release.

The VANDELS passive sample was selected by the following process. Firstly, objects were required to have $1.0 < z_\mathrm{phot} < 2.5$. Objects were then selected to have observed $H$-band magnitudes of $H < 22.5$, corresponding to stellar masses of $\mathrm{log(}M_*/ \mathrm{M_\odot}) \gtrsim 10$. Next, objects were selected by rest-frame UVJ colours (e.g. \citealt{Williams2009}; \citealt{Whitaker2011}). In order to ensure that all targets would be detected with sufficient SNR in the VIMOS spectra, a final selection criterion was applied, requiring an observed $i$-band magnitude of $i < 25$. This slightly biases the full VANDELS passive sample against the faintest and reddest objects. This process, summarised below, results in a parent sample of 812 objects.

\begin{itemize}
\item $1.0 < z_\mathrm{phot} < 2.5$
\smallskip
\item $H < 22.5$
\smallskip
\item $U - V$ > 0.88($V - J$) + 0.49
\smallskip
\item $U - V$ > 1.2
\smallskip
\item $V - J$ < 1.6
\smallskip
\item $i$ < 25.
\end{itemize}

\subsection{VANDELS spectroscopic observations}\label{subsect:data_spectroscopy}

The VANDELS observations are described in full in \cite{Pentericci2018}, and so we again provide here only a brief summary of the relevant points. Of the 812 objects selected by the process detailed in Section \ref{subsect:data_photometry}, a random sample of 268 were assigned slits and observed. All observations were conducted using the MR grism, providing $R\simeq600$ spectroscopy spanning an observed wavelength range from $\lambda = 4800{-}10000$\AA. Objects were observed for either 20, 40 or 80 hours, depending on their $i$-band magnitudes, to obtain SNRs of $15{-}20$ per resolution element ($\sim$10\AA) in the $i$-band. Spectroscopic redshifts, $z_\mathrm{spec}$, were measured and verified by the VANDELS team, as described in section 5 of \cite{Pentericci2018}.

A known issue with the VANDELS spectra is a systematic drop in flux at the blue end ($\lambda \lesssim 5600$\AA). This region typically has a low SNR in the passive spectra, but an empirical correction was derived and implemented based on the bluer star-forming sample (see section 4.1 of \citealt{Pentericci2018}). This median correction is applied to all of the VANDELS spectra, however, object-to-object variations persist at levels of up to $\sim30$ per cent. This calibration uncertainty is fitted as part of our model, as discussed in Section \ref{sect:fitting}.

Because of the rapid build-up of the red sequence across our target redshift range, the passive sample is heavily weighted towards lower redshifts, with 88 per cent having $z_\mathrm{phot} < 1.5$. This means that, for the vast majority of the sample, these spectra contain a full suite of rest-frame UV-optical absorption features, including Mg\textsubscript{UV}, Ca H and K, the 2640\AA, 2900\AA\ and 4000\AA\ breaks and H$\delta$ and higher order Balmer lines, as well as the [O\,\textsc{ii}] 3727\AA\ emission line.

\subsection{The 1.0 < z < 1.3 mass complete sample}\label{subsect:data_sample}

As discussed in Section \ref{subsect:data_photometry}, the VANDELS passive sample is not mass complete across the whole redshift range from $1.0 < z < 2.5$. Furthermore, the full suite of rest-frame UV spectral features described in Section \ref{subsect:data_spectroscopy} is only available at the lower end of this redshift range. Based on these considerations, it was decided to impose additional redshift and stellar-mass limits on the sample to define a mass complete sample for which all of these features are available.

We first apply a limit of $z_\mathrm{spec} < 1.3$, such that the 4000\AA\ break falls blue-ward of the strong sky-line contamination long-ward of 9250\AA. We then return to the v1.0 photometry and derived parameter catalogues and re-apply the initial VANDELS passive sample selection criteria, excepting the $i$-band magnitude limit (see Section \ref{subsect:data_photometry}). For our reduced redshift range, 98 per cent of objects with stellar masses of $\mathrm{log(}M_*/ \mathrm{M_\odot}) > 10.3$ meet the $i$-band limit imposed in Section \ref{subsect:data_photometry}. 

We therefore impose this mass limit, meaning our final sample is a random draw from a 98 per cent mass complete sample. We finally require that objects have received $> 90$ per cent of their final exposure time in VANDELS DR2, and have a spectroscopic redshift quality flag of 3 or 4 (corresponding to > 95 per cent probability of being correct), resulting in a sample of 75 objects. The final sample has a near-uniform distribution in redshift within our chosen limits. Due to the VANDELS observing strategy, our spectra all have similar SNRs. The distribution of 7500\AA\ SNRs is approximately Gaussian, with a mean value of 12.5\AA$^{-1}$ and standard deviation of 5.3\AA$^{-1}$. Key information for each of the objects in our sample is provided in Table \ref{table:objects}.

\begin{table*}
  \caption{Parameters and priors for the model we fit to our data. The first nine parameters are related to our physical model, introduced in Section \ref{sect:model}. The final six are related to our systematic uncertainties model, introduced in Section \ref{subsect:fitting_systematics}. For Gaussian priors, $\mu$ is the mean and $\sigma$ the standard deviation of the prior distribution. The upper limit on the $\tau$ parameter, $t_\mathrm{obs}$, is the age of the Universe at $z_\mathrm{spec}$. Logarithmic priors are uniform in log base ten of the parameter. Our SFH model is a double power law; our calibration model is a second order Chebyshev polynomial. The form of our Gaussian process noise model is given in Equation \ref{eqn:covariance}.}
\begingroup
\setlength{\tabcolsep}{10pt} 
\renewcommand{\arraystretch}{1.1} 
\begin{tabular}{lllllll}
\hline
Component & Parameter & Symbol / Unit & Range & Prior & \multicolumn{2}{l}{Hyperparameters} \\
\hline
Global & Redshift & $z$ & $z_\mathrm{spec} \pm 0.015$ & Gaussian &  $\mu = z_\mathrm{spec}$ & $\sigma$ = 0.005 \\
  & Velocity Dispersion & $\sigma_\mathrm{vel}$ / km s$^{-1}$ & (40, 400) & logarithmic & & \\
\hline
SFH & Stellar mass formed & $M_*\ /\ \mathrm{M_\odot}$ & (1, $10^{13}$) &logarithmic & & \\
 & Metallicity & $Z\ /\ \mathrm{Z_\odot}$ & (0.01, 2.5) &logarithmic & & \\
 & Falling slope & $\alpha$ & (0.1, 1000) & logarithmic & & \\
 & Rising slope & $\beta$ & (0.1, 1000) & logarithmic & & \\
 & Peak time & $\tau$ / Gyr & (0.1, $t_\mathrm{obs}$) & uniform & & \\
 \hline
Dust & Attenuation at 5500\AA\ & $A_V$ / mag & (0, 8) & uniform & & \\
 & Power-law slope & $n$ & (0.3, 1.5) & Gaussian & $\mu = 0.7$ & $\sigma$ = 0.3 \\
\hline
Calibration & Zero order & $P_0$ & (0.5, 1.5) & Gaussian & $\mu = 1$ & $\sigma$ = 0.25 \\
& First order & $P_1$ & ($-0.5$, 0.5) & Gaussian & $\mu = 0$ & $\sigma$ = 0.25\\
& Second order & $P_2$ & ($-0.5$, 0.5) & Gaussian & $\mu = 0$ & $\sigma$ = 0.25\\
\hline
Noise & White noise scaling & $a$ & (0.1, 10)  & logarithmic & & \\
 & Correlated noise amplitude & $b$ / $f_\mathrm{max}$ & (0.0001, 1)  & logarithmic & & \\
 & Correlation length & $l$ / $\Delta\lambda$  & (0.01, 1)  & logarithmic & & \\
\hline
\end{tabular}
\endgroup
\label{table:params}
\end{table*}

\section{Physical model}\label{sect:model}

This section describes the physical model we construct within \bagpipes\ to describe our targets. \bagpipes\ provides a \textsc{Python} framework for self-consistently modelling the stellar, nebular, dust and intergalactic medium properties of galaxy spectra. Nebular emission is modelled by post-processing stellar templates through the \textsc{Cloudy} photoionization code \citep{Ferland2017}, using a method based on that of \cite{Byler2017}. \bagpipes\ also provides several optional dust-attenuation prescriptions, and assumes energy balance, with attenuated light re-radiated in the infrared assuming the models of \cite{Draine2007}. Full details are provided in section 3.1 of \cite{Carnall2018}.

For all of our objects, we vary the observed redshift within a narrow range centred on the spectroscopic redshift, measured as described in section 5 of \cite{Pentericci2018}. We impose a Gaussian prior centred on $z_\mathrm{spec}$, with standard deviation, $\sigma=0.05$, and allow deviations of up to $3\sigma$ in either direction. Velocity dispersion is modelled by convolution of the spectral model with a Gaussian kernel in velocity space. We apply a logarithmic prior to velocity dispersion, $\sigma_\mathrm{vel}$, between 40 and 400 km s$^{-1}$. A summary of the parameters and priors of our model is provided in Table \ref{table:params}. The fitting of  this model to our data is described in Section \ref{sect:fitting}.

\subsection{Stellar population model}\label{subsect:model_stellar}

In this work we use the default \bagpipes\ stellar population models, which are the 2016 updated version of the \cite{bruzual2003} models\footnote{\url{https://www.bruzual.org/~gbruzual/bc03/Updated_version_2016}} (see \citealt{Chevallard2016}). These models have been updated to include the MILES stellar spectral library \citep{Falcon-Barroso2011}, providing $\sim2.5$\AA\ resolution from $3525{-}7500$\AA, and updated stellar evolutionary tracks (\citealt{Bressan2012}; \citealt{Marigo2013}).

We model the chemical-enrichment histories of our galaxies with a delta-function, assuming that all stars within the galaxy have the same metal content with scaled-Solar abundances. This single metallicity is varied with a logarithmic prior between $-2 < \mathrm{log_{10}}(Z_*/\mathrm{Z_\odot}) < 0.4$ (we define Solar metallicity, $\mathrm{Z_\odot} = 0.02$).

We parameterise the SFHs of our galaxies using the double-power-law model described by \cite{Carnall2018}. The SFH is described by 
\begin{equation}\label{eqn:DPL}
\mathrm{SFR}(t)\ \propto\ \Bigg[\bigg(\frac{t}{\tau}\bigg)^{\alpha} + \bigg(\frac{t}{\tau}\bigg)^{-\beta}\Bigg]^{-1}
\end{equation}

\noindent where $\alpha$ is the falling slope, $\beta$ is the rising slope and $\tau$ is related to the peak time. In  \cite{Carnall2018} we demonstrate that this model produces unbiased estimates of the redshifts of formation and quenching for quiescent galaxies from the \textsc{Mufasa} simulation across a wide redshift range.

\subsection{Dust attenuation model}\label{subsect:model_dust}

Recent studies at high redshift have favoured an average attenuation curve slope of $n \simeq 0.7$, where $A_\lambda \propto \lambda^{-n}$ (e.g. \citealt{Cullen2017, Cullen2018}; \citealt{McLure2018a}), similar to that found by \cite{Calzetti2000} for local galaxies. However, several studies suggest significant object-by-object variation (e.g. \citealt{Kriek2013}; \citealt{Narayanan2018}). We therefore model dust attenuation with the modified \cite{Charlot2000} model described in section 3.1.4 of \cite{Carnall2018}. We place a Gaussian prior on $n$, with mean, $\mu=0.7$ and standard deviation, $\sigma=0.3$. We set permissive lower and upper limits of 0.3 and 1.5 respectively. We likewise allow a wide range of $V$-band attenuations, $A_V$, for stellar continuum emission, adopting a uniform prior from $0 < A_V < 8$. We adopt a fixed value of $\epsilon=2$ for the ratio of attenuation between stellar birth clouds and the wider interstellar medium (ISM). We also adopt a value of 10 Myr for $t_\mathrm{BC}$, the lifetime of stellar birth clouds, meaning that $A_V$ is doubled for emission from stars formed in the last 10 Myr.

\subsection{Nebular emission model}\label{subsect:model_nebular}

Our UVJ selection is designed to identify galaxies with low levels of ongoing star-formation, and hence nebular emission. However, it is important for our model to be capable of reproducing the spectra of dusty star-forming galaxies which can contaminate UVJ-selected samples. We therefore implement the nebular emission model described in section 3.1.3 of \cite{Carnall2018}, with a fixed ionization parameter of $\mathrm{log}_{10}(U) = -3$. Light from stars which formed more recently than the lifetime we assume for stellar birth clouds ($t_\mathrm{BC} = 10$ Myr) is processed through our nebular model. The resulting nebular continuum and line emission is attenuated by twice the ISM $A_V$, as described in Section \ref{subsect:model_dust}.

\section{Combining our spectroscopic and photometric data}\label{sect:fitting}

This section describes the fitting of the physical model described in Section \ref{sect:model} to the combined datasets described in Section \ref{sect:data}. We begin in Section \ref{subsect:fitting_history} by reviewing the literature on galaxy spectral fitting. Then, in Section \ref{subsect:fitting_simple}, we show that a simple approach, which does not allow for systematic uncertainties, fails to describe our data. In Section \ref{subsect:fitting_systematics} we construct a model for these systematics, and in Section \ref{subsect:fitting_final} we fit our combined physical plus systematics model to our joint datasets.

\subsection{Historical approaches to spectral fitting}\label{subsect:fitting_history}

Historically, galaxy spectral fitting techniques have been applied to photometric data (e.g. \citealt{Faber1972}), and this has remained a popular approach for several reasons. Photometric data are widely available across a wide range of wavelengths, and both random and systematic uncertainties are relatively simple to characterise (e.g. \citealt{Mortlock2017}; \citealt{McLure2018b}). It is also far simpler to construct models for the broad-band colours of galaxies than for detailed spectral features (e.g. \citealt{Bell2001}; \citealt{Bell2003}), and this has been shown to be sufficient for estimating basic physical properties such as stellar masses (e.g. \citealt{Mobasher2015}).

Such analyses typically assume that the uncertainties on photometric fluxes are well-determined, Gaussian distributed and independent. In this case, the scatter of observed fluxes $f_i$ about their true values follows a chi-squared distribution (where $i$ runs over a number, $N_\mathrm{bands}$, of photometric bandpasses). A physical model which is a function of some parameters, $\Theta$, can therefore be fitted to these observations using the log-likelihood function
\begin{equation}\label{eqn:lnlike_phot}
\mathrm{ln}(\mathcal{L}_\mathrm{phot})\ =\ \ K -0.5\sum_i^\mathrm{N_\mathrm{bands}}\bigg(\dfrac{f_i\ -\ m_i(\Theta)}{\sigma_i}\bigg)^2
\end{equation}

\noindent where $K$ is a constant, $m_i(\Theta)$ is the model prediction for the observed flux $f_i$, and $\sigma_i$ is the corresponding uncertainty.

Significant failures of the above assumptions are fairly simple to identify by assessing the quality of fit, typically using the minimum reduced chi-squared value, and are rare enough that the affected objects can simply be excluded from the analysis. More-subtle failures can be modelled, for example by applying variable zero-point offsets to each band (e.g. \citealt{Brammer2008}), or by asserting that the uncertainty be greater than some fixed fraction of the observed flux (typically 5 per cent; e.g. \citealt{Muzzin2013}; \citealt{Belli2018}). This prevents uncertainties from being underestimated in the high-SNR regime, where the precision of the photometric calibration dominates the error budget.

However, as described in Section \ref{sect:intro}, photometric data are limited in their ability to constrain more-subtle galaxy physical parameters. It has been shown that spectroscopic observations have the potential to improve this situation, however accurate spectrophotometric calibration is notoriously challenging, owing to the need to correct for a range of atmospheric and instrumental effects, such as differential atmospheric refraction, telluric contamination and characterisation of the sensitivity function of the detector. Even for comparatively well-calibrated spectra, wavelength-dependent uncertainties are known to exist, typically at levels of $\sim10$ per cent (e.g. \citealt{Mohler2014}; \citealt{xiang2015}; \citealt{Yan2016}). Historically therefore, analyses of spectroscopic data have been limited to individual spectral features (commonly Lick indices), such that results are independent of spectrophotometric calibration (e.g. \citealt{Faber1985}; \citealt{Gorgas1993}; \citealt{Worthey1994}). Whilst these analyses have produced many extremely valuable results, they do not make use of the full information content of spectroscopic data (e.g. \citealt{Conroy2018}). 

More recently, with the advent of large, well calibrated spectroscopic surveys such as the Sloan Digital Sky Survey (SDSS; \citealt{York2000}), attention has shifted towards full-spectral-fitting methods, which attempt to model and fit the whole information content of spectroscopic data (e.g. \citealt{Heavens2000, Heavens2004}; \citealt{Panter2003, Panter2007}; \citealt{CidFernandes2005}; \citealt{Ocvirk2006}; \citealt{Tojeiro2007}). The simplest approach to full spectral fitting is to make the same assumption of well-determined, independent Gaussian uncertainties on each spectral pixel flux $f_j$ (where $j$ runs over the number of pixels in the spectrum, $N_\mathrm{pix}$). In this case, the log-likelihood function can again be written as
\begin{equation}\label{eqn:lnlike_spec}
\mathrm{ln}(\mathcal{L}_\mathrm{spec})\ =\ \ K -0.5\sum_j^\mathrm{N_\mathrm{pix}}\bigg(\dfrac{y_j\ -\ m_j(\Theta)}{\sigma_j}\bigg)^2
\end{equation}

\noindent where $K$ is a constant, $m_j(\Theta)$ is the model prediction for the pixel flux $f_j$, and $\sigma_j$ is the corresponding uncertainty. 

This approach has however been demonstrated to be less successful in describing spectroscopic data than photometry. \cite{Panter2003} note that the quality of the fits they obtain are typically poor, and attribute this to both inadequacies in the models they fit to their data, and the difficulty of obtaining reliable uncertainties. Furthermore, in \cite{Panter2007}, the authors report that improvements to the SDSS spectrophotometric calibration have significantly changed their inferred SFHs. These and other authors also note the challenges which exist in the exploration of the higher-dimensional parameter spaces of the more complex models required to fit spectroscopic data. 

As the availability of high-quality panchromatic photometric data has increased, interest has grown in the use of photometry to complement spectroscopic analyses. These joint analyses promise improvements in our understanding by taking advantage of both the broad wavelength coverage and excellent calibration of photometry, and the strong constraints on subtle physical parameters offered by spectroscopy (e.g. \citealt{Chevallard2016}; \citealt{Belli2018}).

\begin{figure*}
	\includegraphics[width=\textwidth]{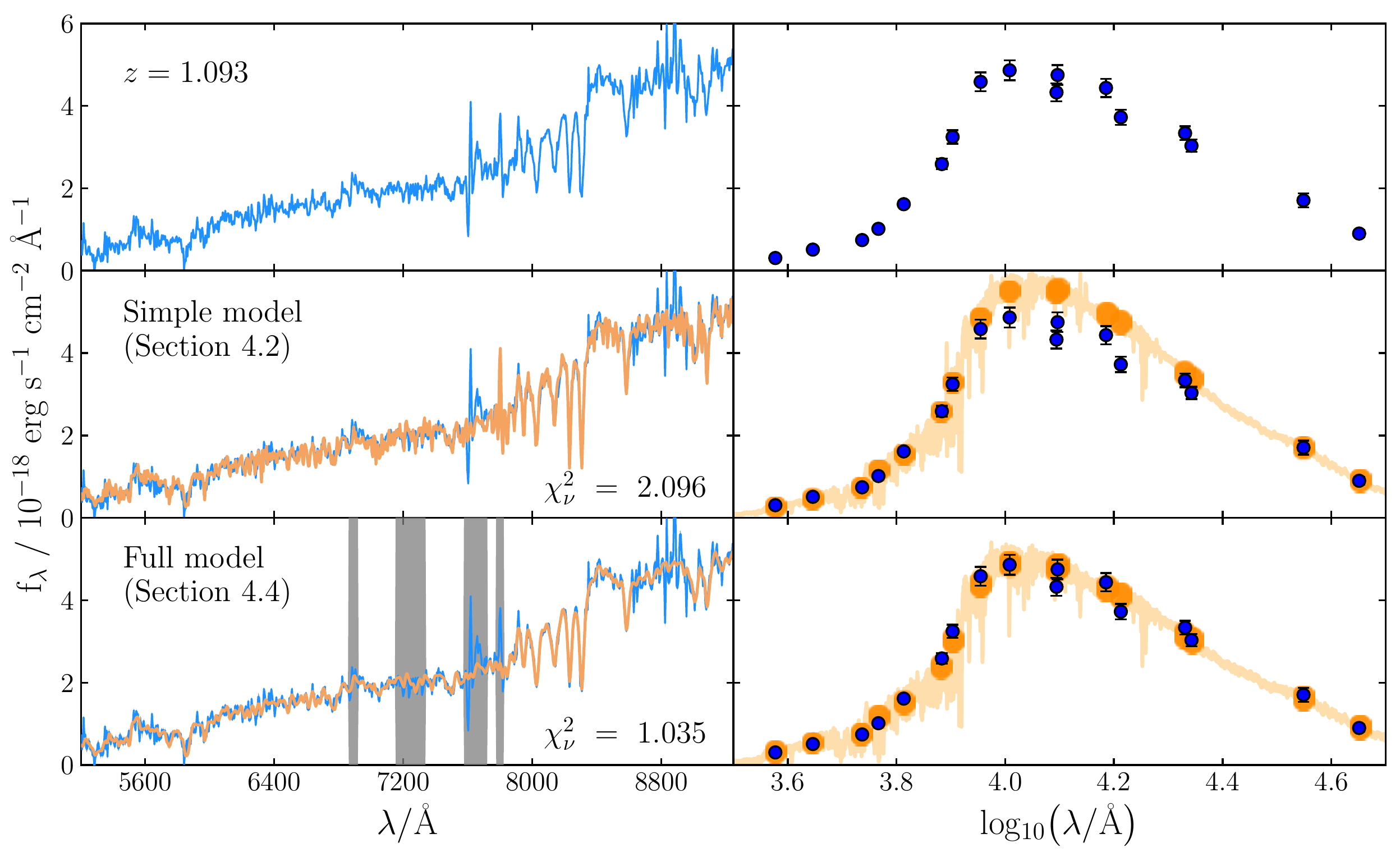}
    \caption{An example object from our sample fitted using different methods. Data are shown in blue, with spectroscopic data to the left and photometric data to the right. Posterior distributions are shown in orange (the $16^\mathrm{th}{-}84^\mathrm{th}$ percentiles) for the simple model of Section \protect \ref{subsect:fitting_simple} in the middle panels and for the final joint physical plus systematics model of Sections \ref{subsect:fitting_systematics} and \ref{subsect:fitting_final} in the bottom panels. The gray shaded regions were  masked in our final fits (see Section \ref{subsect:fitting_final}). Minimum reduced chi-squared values for both fits, $\chi^2_\nu$, are also shown. The $\chi^2_\nu$ value for the simple model has been divided through by the maximum likelihood value of $a^2$ from the full model fit, such that the same spectroscopic uncertainties are assumed in both cases. An expanded version of the bottom-left panel is shown in Fig. \protect \ref{fig:fit_systematics}.}\label{fig:fit_comparison}
\end{figure*}

\subsection{A simple approach to fitting our joint datasets}\label{subsect:fitting_simple}

When jointly analysing several datasets, the likelihood function is the product of the separate likelihoods. The log-likelihood function for the joint analysis of spectroscopy and photometry is therefore $\mathrm{ln}(\mathcal{L}) =  \mathrm{ln}(\mathcal{L}_\mathrm{phot})\ +\ \mathrm{ln}(\mathcal{L}_\mathrm{spec})$. In this section we take a simple approach, by attempting to jointly fit our spectroscopic and photometric datasets using the physical model described in Section \ref{sect:model}, and the ansatz for $\mathrm{ln}(\mathcal{L}_\mathrm{phot})$ and $\mathrm{ln}(\mathcal{L}_\mathrm{spec})$ given in Equations \ref{eqn:lnlike_phot} and \ref{eqn:lnlike_spec} respectively. This approach will henceforth be described as our simple model. The nine parameters of this model are summarised at the start of Table \ref{table:params}. The other six parameters below are only included in the final model, which is described in Sections \ref{subsect:fitting_systematics} and \ref{subsect:fitting_final}.

For our photometric data we employ the common methods discussed in Section \ref{subsect:fitting_history} for dealing with systematic calibration uncertainties. We first apply the photometric zero-point offsets calculated for the VANDELS photometric catalogues by \cite{McLure2018b}. We also assert that the uncertainty for each band must be $\geq 5$ per cent of the observed flux, except for the two IRAC channels where a threshold of 10 per cent is used. The VANDELS DR2 spectra were binned by a factor of two to a sampling of 5\AA. Fitting was carried out using the wavelength range from $5200$\AA\ $< \lambda < 9250$\AA\ where the detector sensitivity is high and sky-line contamination is minimal compared to longer wavelengths. Sampling of the posterior was carried out using the \textsc{MultiNest} algorithm \citep{Feroz2008, Feroz2009, Feroz2013}.

The results of this method applied to an example object from our sample are shown in the middle panels of Fig. \ref{fig:fit_comparison}. It can be seen that the overall shape of the posterior distribution matches the spectroscopic data well. However, on closer inspection (and comparison to the above panel), it can be seen that the depths of individual absorption features are poorly reproduced. The most obvious failure however is in reproducing the observed photometry from $1{-}2$ $\mu$m, where our model significantly overestimates the observed fluxes. The object shown in Fig. \ref{fig:fit_comparison} is typical of the sample, and these issues were observed for the majority of our galaxies.

The fact that the joint fit is incapable of matching both datasets simultaneously means that the two are inconsistent under the assumptions we have made. The joint fit adopts a region of parameter space which best describes the spectroscopic data at the expense of the photometry because there are a larger number of spectral pixels than photometric bands, and hence more terms in the log-likelihood function which depend on the spectroscopy than the photometry. Even though the joint fit is dominated by the spectroscopy, the quality of fit to the spectroscopic data is still poor, in accordance with the results of \cite{Panter2003}.

Issues of this nature have been commonly observed in similar analyses, leading to suggestions that the spectroscopic data should be somehow down-weighted in the likelihood function, in order to give ``equal consideration'' to both datasets. However, apart from being statistically unjustified, this cannot solve the underlying issues of inconsistency between the datasets and poor quality of fit to the spectroscopic data. Instead it is necessary to understand the causes of these issues, so that these effects can be included in the model we fit to our joint datasets.

One possible reason for inconsistencies between our datasets would be aperture bias, as our photometry is measured within $2^{\prime\prime}-$diameter apertures, whilst our spectroscopic observations use a $1^{\prime\prime}$ slit. However, similar studies at lower redshifts have found aperture effects to be small (e.g. \citealt{Gallazzi2014}), and we would expect a smaller effect still, due to the larger angular diameter distance to our target redshift range and the smaller physical sizes of quiescent galaxies at higher redshifts (e.g. \citealt{McLure2013}). As a check, we take effective radius measurements, $r_\mathrm{e}$, from \cite{vanderwel2014} for the 26 objects which have CANDELS imaging. The mean $r_\mathrm{e}$ is $0.35^{\prime\prime}$, and no correlation exists between  $r_\mathrm{e}$ and the implied inconsistency between our datasets. Aperture bias also cannot explain the poor quality of fit to the spectroscopic data.

As described in Section \ref{subsect:data_spectroscopy}, there is a known issue with the VANDELS spectrophotometric calibration, which is typically low at the blue end, reddening the spectra with respect to the observed photometry. As the joint fits are dominated by the spectroscopic data, this would be expected to cause the fitted models to be redder than the observed photometry, which is consistent with the overestimation of the $1{-}2$ $\mu$m photometry observed in the middle panels of Fig.~\ref{fig:fit_comparison}. Incorrect spectroscopic calibration is also likely to lead to a poor quality of fit, as it is unlikely that models will exist within the parameter space being explored which can simultaneously match the perturbed spectral shape and absorption features present. We therefore conclude that systematic calibration uncertainties in our spectroscopic data are the most likely cause of the issues we identify, and move on to develop a model for these systematic uncertainties.

\subsection{Modelling spectroscopic systematic uncertainties}\label{subsect:fitting_systematics}

In general, as discussed in Section \ref{subsect:fitting_history}, it is extremely complex to construct a physical model for the atmospheric and instrumental effects to which both our spectroscopic data and our empirical stellar-population models are subjected. We therefore take a different approach, by constructing a flexible empirical model for systematic perturbations of our spectroscopic data about our physical model. 

We will split these perturbations by the general form they take: either additive or multiplicative (e.g. \citealt{Cappellari2017}). We will refer to these as noise and calibration offsets respectively. We then construct flexible models for these offsets by introducing nuisance parameters, $\Phi$, into the spectroscopic log-likelihood function. We can then later marginalise these nuisance parameters out of our posterior distribution, in order to obtain a posterior for our physical parameters which includes uncertainties due to systematic effects.

We modify the log-likelihood function for spectroscopy presented in Equation \ref{eqn:lnlike_spec} as follows. Firstly we generalise our model, $m(\Theta)$ to  $m(\Theta, \Phi)$ by dividing through by a multiplicative polynomial calibration model, $P_j(\Phi)$, such that 
\begin{equation}\label{eqn:polynomial}
m_j(\Theta,\ \Phi) = \dfrac{m_j(\Theta)}{P_j(\Phi)}.
\end{equation}

\noindent This model will be discussed in Section \ref{subsubsect:fitting_polynomial}.

Secondly, we drop the assumption that the uncertainties on our observed spectroscopic fluxes are independent, allowing additive correlated noise between our spectral pixels. We hence replace the second term on the right of Equation \ref{eqn:lnlike_spec} with a matrix equation, in which the inverse of the covariance matrix is multiplied on both sides by the vector of residuals between our observed and model fluxes. Our covariance matrix, $\mathbf{C}(\Phi)$, will be drawn from a Gaussian process model, which is described in Section \ref{subsubsect:fitting_noise}. Our spectroscopic log-likelihood function is now
\begin{equation}\label{eqn:lnlike_spec_new}
\mathrm{ln}(\mathcal{L}_\mathrm{spec})\ =\ \ K - \mathrm{ln}\big(|\mathbf{C}(\Phi)|\big)\ - \Delta^\mathrm{T}\ \mathbf{C}(\Phi)^{-1}\ \Delta
\end{equation}

\noindent where $\Delta\ =\ y_j\ -\ m_j(\Theta,\ \Phi)$ is the vector of residuals between our observed and model fluxes and $K$ is a constant. Equation \ref{eqn:lnlike_spec_new} is simply a generalisation of Equation \ref{eqn:lnlike_spec}, and reduces back to Equation \ref{eqn:lnlike_spec} for the case in which $P_j(\Phi) = 1$ and $\mathbf{C}(\Phi)$ is diagonal with elements $\sigma_j^2$.

\begin{figure*}
	\includegraphics[width=\textwidth]{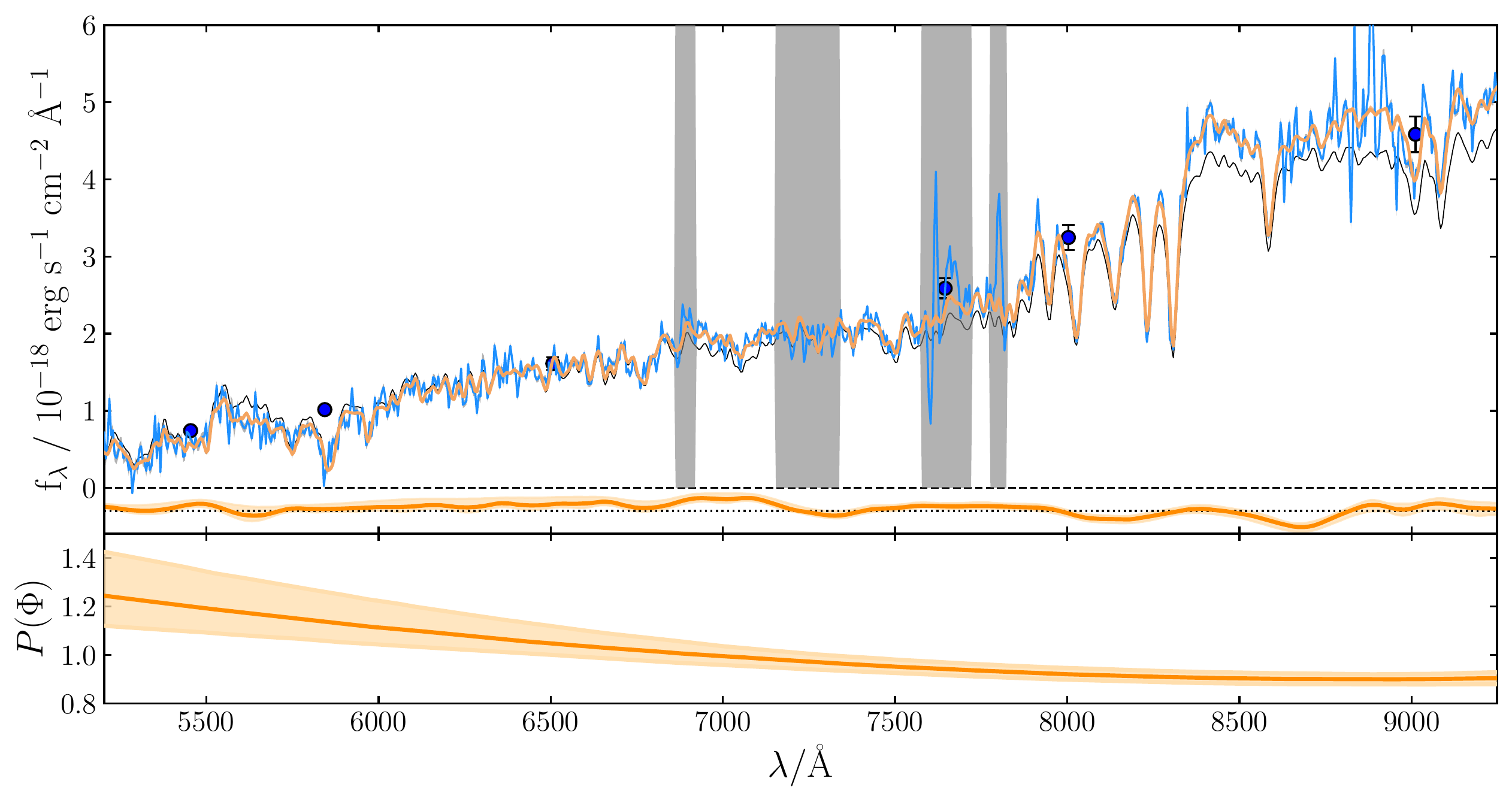}
    \caption{Physical, noise and calibration models fitted to the same example object as shown in Fig. \ref{fig:fit_comparison}. The observed spectrum is shown in blue. The gray shaded regions have been masked, as described in Section \ref{subsect:fitting_final}. The orange line overlaid on the spectrum shows the posterior distribution for the final physical + calibration + noise model. The posterior distribution for the Gaussian process noise model is shown below in the same panel at an arbitrary vertical position. The zero-point about which it varies is shown with a dotted line. The posterior distribution for the polynomial calibration model is shown in the bottom panel. The posterior median for the physical model alone is shown in black. This is analogous to the spectrum corrected for systematic effects. To obtain the median of the orange posterior in the top panel, the black line is divided by the polynomial at the bottom, then added to the Gaussian process model in the centre.}\label{fig:fit_systematics}
\end{figure*}

\subsubsection{Polynomial calibration model}\label{subsubsect:fitting_polynomial}

A method recently adopted by several authors for addressing spectrophotometric calibration uncertainties is to perturb the spectroscopic data by a polynomial function of wavelength. One approach is to set the polynomial coefficients before fitting a physical model to the data, by comparing synthetic photometry derived from the spectrum to observed photometry in the same wavelength range (e.g. \citealt{vanderwel2016}). This  is computationally simple, however it requires a significant number of photometric observations in the spectroscopic wavelength range. Another approach is to fit the polynomial coefficients at the same time as the parameters of the physical model, which has the advantage of incorporating calibration uncertainties into the uncertainties on physical parameters, but is more computationally expensive (e.g. \citealt{Cappellari2017}; \citealt{Belli2018}).

We take the latter approach, parameterising $P_j(\Phi)$ in Equation \ref{eqn:polynomial} with a second-order Chebyshev polynomial. Because our physical model is divided by $P_j(\Phi)$, the polynomial posterior can be thought of as the multiplicative offset which would need to be applied to our data to correct its calibration. We caution however that any issues with the spectrophotometric calibration of our model spectra will also be incorporated into this polynomial. 

The choice of a second order polynomial was made firstly due to computational constraints, and secondly because the calibration issues with the VANDELS spectra are known to be well approximated by a quadratic function of wavelength (see fig. 4 of \citealt{Pentericci2018}). A more flexible choice for this model would be a multiplicative Gaussian process model (e.g. Johnson et al. in prep), and we intend to explore this option in future work.

We apply Gaussian priors to all three polynomial coefficients with standard deviations of $\sigma = 0.25$. The prior means are $\mu=1$ for the zero order, and $\mu=0$ for the first and second order. This means that the prior mean and median for $P_j(\Phi)$ are equal to 1 for all wavelengths, equivalent to no change in the calibration. The maximum deviation allowed for any polynomial order is $2\sigma$ from the prior mean.

\subsubsection{Gaussian process noise model}\label{subsubsect:fitting_noise}

As discussed in Section \ref{subsect:fitting_history}, it is common to assume that uncertainties on observational data are independently Gaussian distributed with well-known variances. There are many good reasons to suspect that these assumptions do not hold in the case of spectroscopic data. 

Firstly, it is standard practice to oversample the resolution element of the optical system by at least a factor of two, leading to local covariances between pixels, although this can be mitigated to some extent by binning adjacent pixels. Secondly, there is also good evidence that the error spectra determined from typical data reduction pipelines are underestimates of the true pixel variances (e.g. \citealt{Panter2003, Panter2007}; \citealt{Belli2018}). A common approach is to expand the variances for all spectral pixels by the median residual determined from an initial round of fitting (e.g. \citealt{Belli2018}). Finally, a range of effects from template mismatch to poor sky subtraction have the potential for introducing correlated additive offsets between the data and models being fitted (e.g. \citealt{Cappellari2017}). 

By modifying the log-likelihood function presented in Equation \ref{eqn:lnlike_spec} to that of Equation \ref{eqn:lnlike_spec_new}, we have relaxed the assumption of independence in our spectroscopic uncertainties. We now parameterise our covariance matrix, $\mathbf{C}(\Phi)$, in terms of both independent (white) noise and covariant noise between pixels. We will fit these parameters alongside those of our physical and polynomial calibration models. The form we assume for our covariance matrix is
\begin{equation}\label{eqn:covariance}
\mathbf{C}_{jk}(\Phi)\ =\ a^2\sigma_j\ \sigma_k\ \delta_{jk}\ +\ b^2\mathrm{exp}\bigg(-\dfrac{(\lambda_j - \lambda_k)^2}{2l^2}\bigg)
\end{equation}

\noindent where $\sigma_{j, k}$ are the uncertainties on our pixel fluxes, $\lambda_{j, k}$ are the central wavelengths of our pixels, $\delta_{jk}$ is the Kronecker delta function, and $a$, $b$ and $l$ are free parameters to be fitted.

The first term in Equation \ref{eqn:covariance} deals with the uncorrelated noise on our data. As we suspect that our uncertainties may be underestimated, we allow their magnitude to vary by $a^2$, where $a$ is assigned a logarithmic prior between 0.1 and 10 (e.g. see section 6 of \citealt{Hogg2010}). This is similar to the iterative approaches of other authors, however the fact that this parameter is allowed to vary during fitting means that its uncertainty is propagated into the uncertainties on our physical parameters.

The second term in Equation \ref{eqn:covariance} is drawn from a Gaussian process model, and allows us to model covariant noise between our spectral pixels. Gaussian process regression is implemented in \bagpipes\ using the \textsc{George} Python package (\citealt{Ambikasaran2015}). We adopt an exponential-squared kernel and fit the normalisation $b$ and correlation length $l$. We assign logarithmic priors to both of these quantities. We define $b$ in units of the maximum flux in the observed spectrum, $f_\mathrm{max}$, and allow values from $10^{-4}$ to 1. The maximum flux is used as the unit of $b$ such that the same range of prior values can be used for each spectrum. The mean or median flux value could also have been used. 

Similarly, we define $l$ in units of the wavelength range covered by the spectral data, $\Delta\lambda$ (in this case 4050\AA), and allow values from 0.01 to 1. The minimum correlation length ($\sim40$\AA) was chosen to prevent the Gaussian process model from reproducing individual absorption and emission features in our spectra. Our Gaussian process model is intended to model poor sky subtraction and template mismatch between our models and data, as demonstrated with an additive polynomial by \cite{Cappellari2017}. As currently implemented, it cannot model covariances between adjacent spectral fluxes due to oversampling of the resolution element, or resampling from an initial non-uniform wavelength sampling. A term in Equation \ref{eqn:covariance} to account for this is a possible extension to our model (e.g. \citealt{Czekala2015}).

\subsection{Final fitting of our joint datasets}\label{subsect:fitting_final} 

In this section we describe our final fitting methodology, from which all of the results presented in Section \ref{sect:results} are derived. We again fit the physical model described in Section \ref{sect:model}, however we now also fit the models for systematic effects introduced in Section \ref{subsect:fitting_systematics}, by exchanging Equation \ref{eqn:lnlike_spec} for Equation \ref{eqn:lnlike_spec_new} in our log-likelihood function. 

Our photometric data are treated in the same way as described in Section \ref{subsect:fitting_simple}, and we use the same wavelength range and binning for our spectroscopy. In addition, we also mask several spectral regions which experience strong telluric contamination, leading to residuals such as those visible in the top left panel of Fig. \ref{fig:fit_comparison} at  $\sim7600$\AA. The regions masked are $6860{-}6920$\AA, $7150{-}7340$\AA, and $7575{-}7725$\AA. Finally, we mask the rest-frame region from $3702{-}3752$\AA, containing the [O\,\textsc{ii}] emission line. This is because the excitation mechanism for low-level line emission in quiescent galaxies is still controversial, with AGN and ionization from old stars both thought to contribute (e.g. \citealt{Yan2006}; \citealt{Lemaux2010}; \citealt{Singh2013}; \citealt{Herpich2018}). By contrast, the only mechanism which can excite [O\,\textsc{ii}] emission in our \bagpipes\ physical model is ionization from young stars, meaning that our inferred SFRs could be biased by [O\,\textsc{ii}] emission excited by other processes. We will compare our observed [O\,\textsc{ii}] equivalent widths to our inferred specific star-formation rates (sSFRs) in Section \ref{subsect:results_oii}.

Our final model has 15 free parameters, summarised in Table \ref{table:params}. Sampling our posterior distribution with \textsc{MultiNest} therefore requires several million evaluations of our log-likelihood function, each of which is relatively computationally expensive, in particular the inversion of the covariance matrix. Fitting each galaxy therefore requires several tens of CPU hours, limiting the scalability of this method.

The posterior distribution for our final model fitted to the object discussed in Section \ref{subsect:fitting_simple} is shown in the bottom panels of Fig. \ref{fig:fit_comparison}. Both the spectroscopic and photometric data can now be seen to be well matched by our posterior distribution. In order to provide a quantitative comparison, we also show in Fig. \ref{fig:fit_comparison} minimum reduced chi-squared values, $\chi_\nu^2$, for both of the fits shown. In order for the comparison to be direct, the $\chi_\nu^2$ for the fit with the simple model has been divided by the maximum likelihood value of $a^2$ from the fit with the full model (see Equation \ref{eqn:covariance}). This means that the same variances are assumed for the spectroscopic data in both cases. When comparing to a chi-squared distribution with 794 degrees of freedom, the full model fit is statistically acceptable (24 per cent chance of a worse fit), whereas the simple model fit is strongly excluded by the data. The $\chi_\nu^2$ values shown in Fig. \ref{fig:fit_comparison} are typical of galaxies in our sample.

An expanded view of the bottom left panel of Fig. \ref{fig:fit_comparison} is shown in Fig. \ref{fig:fit_systematics}. Additionally, the posterior distributions for our polynomial calibration and Gaussian process noise models are shown below the observed spectrum. The posterior distribution for our polynomial calibration model can be seen to follow the expected form, with little change across most of the spectral range, and an increase of $\sim30$ per cent at the blue end. This is typical of the polynomial corrections we recover for objects in our sample, although the degree of correction at the blue end varies by an additional $\sim30$ per cent from object to object. These calibration corrections follow the expected form found by \cite{Pentericci2018}, and eliminate the apparent inconsistencies between spectroscopic and photometric datasets discussed in Section \ref{subsect:fitting_simple}. The corrections introduced by our Gaussian process noise model can be seen to be small, however the extra flexibility in continuum shape allows the absorption features present in the observed spectrum to be well fitted by our model.

Finally, the black line in the main panel of Fig. \ref{fig:fit_systematics} shows the posterior median for the physical model we fit to our observed spectrum. Assuming that systematics on our spectral models are negligible, this can be thought of as the best fit to our observational data corrected for systematic effects. For clarity, the black line in the top panel divided by the polynomial in the bottom panel then added to the Gaussian process model in the centre gives the median of the orange posterior distribution shown in the top panel.

\section{Results}\label{sect:results}

In this section we present the results obtained by fitting our full model, as described in Sections \ref{subsect:fitting_systematics} and \ref{subsect:fitting_final}, to the sample described in Section \ref{subsect:data_sample}.

\subsection{Quiescent and green valley sub-samples}\label{subsect:results_classification}

Two main methods have been used to define samples of quiescent galaxies: selection by sSFR and selection by rest-frame UVJ colours (typically evolving with observed redshift; e.g. \citealt{Williams2009}). Several recent studies define quiescence by a time-evolving criterion of sSFR $< 0.2/t_\mathrm{obs}$, where $t_\mathrm{obs}$ is the age of the Universe when the galaxy is observed (e.g. \citealt{Gallazzi2014}; \citealt{Pacifici2016}). This was demonstrated by \cite{Carnall2018} to produce good agreement with UVJ selection using a non-evolving colour criterion of $U - V$ > 0.88($V - J$) + 0.69 from $0.25 < z < 3.75$.

As detailed in Section \ref{sect:data}, the VANDELS UVJ selection uses the more permissive $1.0 < z < 2.0$ colour criterion from \cite{Williams2009} of $U - V$ > 0.88($V - J$) + 0.49. We therefore apply a further, sSFR-based selection to our sample in order to facilitate comparisons with other recent work. We apply the slightly modified method of \cite{Carnall2018}, which uses the normalised SFR, or nSFR. This is defined as the SFR averaged over the most recent 100 Myr, SFR$_{100}$, as a fraction of the average SFR over the whole history of the galaxy. This can be written as
\begin{equation}\label{eqn:nsfr}
\mathrm{nSFR}\ =\ \mathrm{SFR_{100}}\dfrac{t_\mathrm{obs}}{\int_{0}^{t_\mathrm{obs}} \mathrm{SFR}(t)\ \mathrm{d}t}.
\end{equation}

\noindent In fig. 5 of \cite{Carnall2018} we demonstrate that a selection criterion of nSFR < 0.1 produces good agreement with sSFR $< 0.2/t_\mathrm{obs}$ at all redshifts. This requires the ongoing SFR of the galaxy at the redshift of observation to be less than 10 per cent of its historical average. By the application of this criterion we separate our sample into 53 quiescent and 22 green-valley galaxies. This approach will be compared to UVJ-based selection criteria in Section \ref{subsubsect:results_uvj_ssfr}.

\begin{figure}
	\includegraphics[width=\columnwidth]{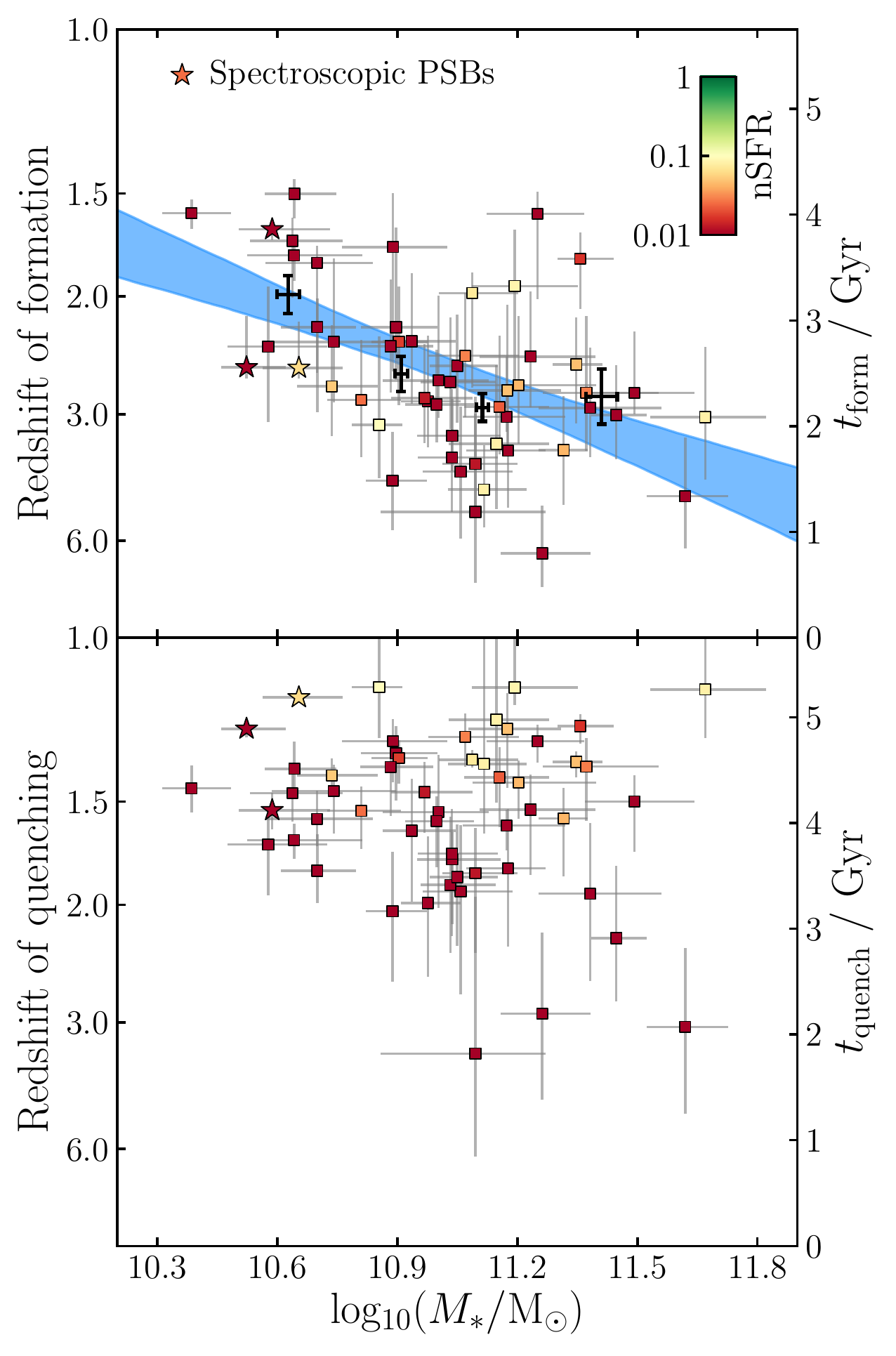}
    \caption{Redshifts of formation (top panel) and quenching (bottom panel) for our quiescent sub-sample. In the top panel the black errorbars show the mean time of formation in four stellar mass bins. The posterior distribution given in Equation \ref{eqn:tform} for the mean formation time as a function of stellar mass is shown in blue. Individual SFHs for these objects are shown in Fig. \ref{fig:sfh_post}.}\label{fig:ages}
\end{figure}

\begin{figure*}
	\includegraphics[width=\textwidth]{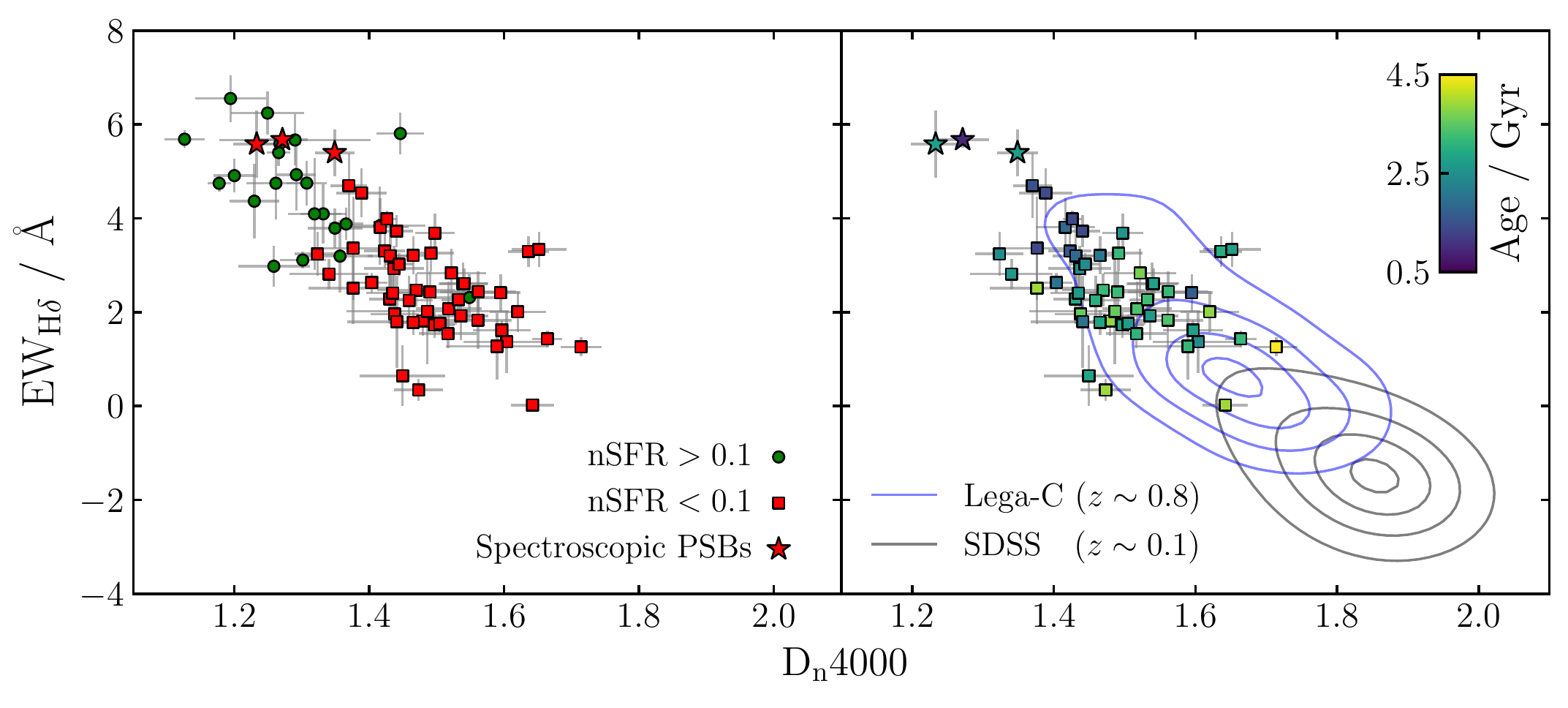}
    \caption{Distribution of our galaxies in EW$_\mathrm{H\delta}$ vs D\textsubscript{n}4000. To the left, our quiescent (red) and green-valley (green) sub-samples are shown, and can be seen to be cleanly separated in this parameter space at D\textsubscript{n}4000 $\sim1.3{-}1.4$ and EW$_{\mathrm{H}\delta}\sim4$\AA. To the right, our quiescent sub-sample is shown coloured by the inferred mass-weighted age. Contours are also shown marking the distributions of SDSS (gray) and Lega-C (blue) quiescent galaxies, also selected by nSFR < 0.1, from the samples of \protect \citet[priv. comm.]{Wu2018a}.}\label{fig:d4000_hd}
\end{figure*}

\subsection{Stellar mass vs formation redshift}\label{subsect:results_age}

From the star-formation histories we infer for our quiescent sub-sample we measure several quantities. Firstly, we calculate the mean time (measured forwards from the Big Bang) at which the stars in our galaxies formed, $t_\mathrm{form}$, given by
\begin{equation}\label{eqn:tform}
t_\mathrm{form} = \frac{\int_{0}^{t_\mathrm{obs}} t\ \mathrm{SFR}(t)\ \mathrm{d}t}{\int_{0}^{t_\mathrm{obs}} \mathrm{SFR}(t)\ \mathrm{d}t}.
\end{equation}

\noindent This corresponds to the mean stellar age, or mass-weighted age, and is similar to the median formation time, $t_{50}$, used in some studies. We then calculate the redshift corresponding to $t_\mathrm{form}$, which we call the formation redshift.

We also calculate the history of the nSFR parameter over our inferred SFHs and extract the age of the Universe at which nSFR first falls below 0.1. This is the time at which the galaxy would first enter our quiescent sub-sample, and we therefore refer to this as the time of quenching, $t_\mathrm{quench}$.

Times of formation and quenching inferred using the double-power-law SFH model described in Section \ref{subsect:model_stellar} were extensively validated using mock photometric observations of simulated quiescent galaxies by \cite{Carnall2018}. However, subsequent evidence (\citealt{carnall2019}; Iyer et al. in prep.) suggests that this SFH model under-predicts the ages of star-forming galaxies. We therefore do not report inferred ages for our green-valley sub-sample.

The times of formation and quenching we infer for our quiescent sub-sample are shown in Fig. \ref{fig:ages} as a function of our inferred stellar masses. It can be seen that almost all of our lower-mass galaxies $(M_* \lesssim 10^{11} \mathrm{M_\odot})$ are found to have formed the bulk of their stellar populations at $z < 3$. By contrast, more-massive galaxies display a greater spread in formation redshift, with the oldest objects found to have formed their stars at $z \sim 5$ and quenched by $z=3$.

In order to understand the average properties of our sample as a function of stellar mass, we split our objects into four mass bins. A 0.25 dex spacing was used between log$_{10}(M_*/\mathrm{M}_\odot) = 10.75$ and 11.25, with additional bins for objects with $M_*~<~10^{10.75}\mathrm{M}_\odot$ and $M_*~>~10^{11.25}\mathrm{M}_\odot$. The mean formation redshift in each bin is shown with black errorbars in the top panel of Fig. \ref{fig:ages}. A clear trend is visible at lower masses, which shows some signs of flattening at $M_* > 10^{11} \mathrm{M_\odot}$, as found at $z=0.7$ by \cite{Gallazzi2014}.

Following \cite{Gallazzi2014} we fit a linear relationship to describe the average $t_\mathrm{form}$ for our galaxies as a function of stellar mass, including an intrinsic scatter to account for effects unrelated to the stellar masses of our objects (e.g. galaxy environment). For the mean relationship we find
\begin{equation}\label{eqn:tform}
\bigg(\dfrac{t_\mathrm{form}}{\mathrm{Gyr}}\bigg)\ =\ 2.56^{+0.12}_{-0.10}\ -\ 1.48^{+0.34}_{-0.39}\ \mathrm{log}_{10}\bigg(\dfrac{M_*}{10^{11}\mathrm{M_\odot}}\bigg)
\end{equation}

\noindent with an intrinsic scatter of $0.58^{+0.09}_{-0.08}$ Gyr. The posterior for the mean relationship is shown in blue in the top panel of Fig. \ref{fig:ages}. This result will be discussed in Section \ref{subsect:discussion_downsizing}. 

It is interesting to consider the distribution of SFRs for the objects shown in Fig. \ref{fig:ages}. For all objects with nSFR < 0.01 (those coloured dark red) the lower bound on the current level of star-formation is zero: we hence describe these objects as having no detectable star-formation. Galaxies with lighter colours have detectable star-formation at levels low enough that we still describe them as quiescent. It can be seen that no trend exists between $t_\mathrm{form}$ and nSFR in Fig. \ref{fig:ages}. This supports the findings of \cite{Belli2017}, who attribute star-formation in quenched galaxies to stochastic processes such as minor mergers and rejuvenation events. 

A trend is visible between $t_\mathrm{quench}$ and nSFR, with more recently quenched galaxies having higher nSFR. However this result is not robust, as our double-power-law SFH model cannot reproduce rejuvenation events. Once star-formation drops to near zero it cannot rise again under this model, meaning quenching must be delayed until recent times to match any level of ongoing star formation. We hence conclude that we cannot reliably measure $t_\mathrm{quench}$ with our double-power-law model from UV spectroscopy for galaxies with detectable ongoing star-formation. A more advanced approach will be needed to model the details of these SFHs (e.g. \citealt{Leja2019a}; \citealt{Iyer2019}; \citealt{Lovell2019}).

\begin{figure*}
	\includegraphics[width=\textwidth]{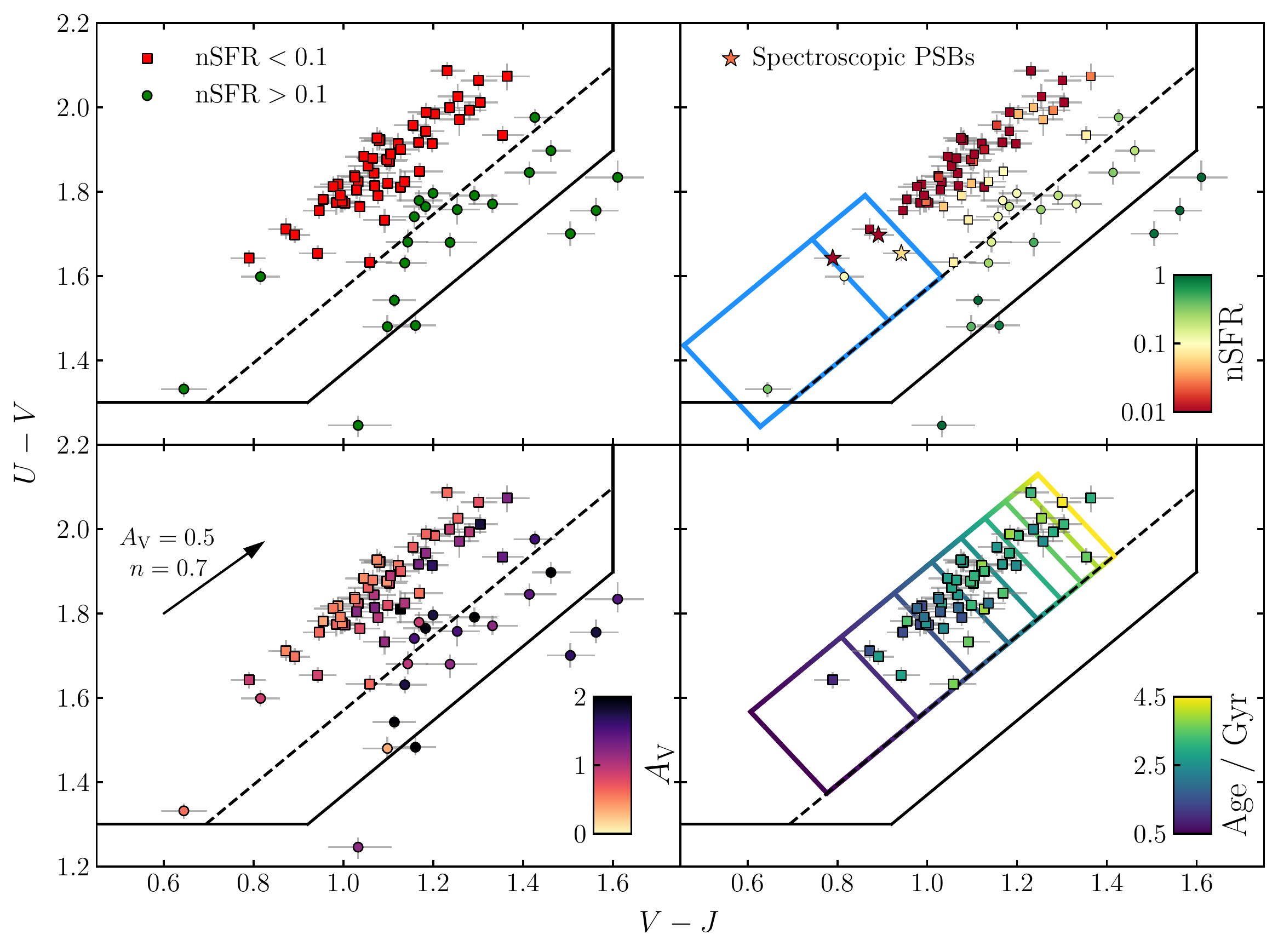}
    \caption{Our sample on the UVJ diagram, coloured by nSFR, $A_\mathrm{V}$ and mass-weighted age. The VANDELS UVJ criteria (Section \ref{subsect:data_photometry}) are shown by solid black lines; the dashed line is the stricter boundary of $U - V$ > 0.88($V - J$) + 0.69. In the top-right panel the larger blue box is the PSB selection of \protect \cite{Belli2018}, assuming that galaxies with median stellar ages of $300{-}800$ Myr display PSB spectral properties. The upper box shows an extension of the PSB selection to a maximum age of 1.2 Gyr. In the bottom-left panel, the arrow shows the effect of 0.5 magnitudes of $V$-band attenuation, under the model described in Section \ref{subsect:model_dust} with $n=0.7$. In the bottom-right panel, the coloured grid shows predicted positions from \protect \cite{Belli2018} for median stellar ages from $0.5{-}4.5$ Gyr in 0.5 Gyr intervals.}\label{fig:uvj}
\end{figure*}

\subsection{Distribution in D\textsubscript{n}4000 vs H$\boldsymbol{\delta}$}\label{subsect:results_d4000_hd}

Historically, a common method for inferring galaxy ages and sSFRs from UV-optical spectroscopy has been to 
measure the strengths of the 4000\AA\ break (D\textsubscript{n}4000) and the Balmer delta (H$\delta$) absorption feature (e.g. \citealt{Kauffmann2003}; \citealt{Brinchmann2004}). We therefore report these spectral indices for our galaxies, both to show the distribution of these parameters within the quiescent population at $1.0 < z < 1.3$, and to check that the results we infer from our full-spectral-fitting method are in agreement with the expected relationships between these parameters. 

We measure D\textsubscript{n}4000 from our spectra as the ratio of average fluxes between $3850{-}3950$\AA\ and $4000{-}4100$\AA, whilst masking out pixels which deviate by more than $3\sigma$ from the posterior median model fitted in Section \ref{subsect:fitting_final}, which are assumed to experience significant sky-line contamination (31 objects have pixels masked, with the average fraction masked being 2 per cent). We measure the rest-frame H$\delta$ equivalent width, EW$_{\mathrm{H}\delta}$, by fitting a first-order polynomial plus Gaussian model to the 100\AA\ spectral region centred on H$\delta$. We then correct for nebular emission using the H$\delta$ flux predicted by our fitted \bagpipes\ model (see Section \ref{subsect:model_nebular}). These corrections are small ($\lesssim1$\AA) for our green valley sub-sample, and typically negligible for our quiescent sub-sample. For 8 of our 75 objects, H$\delta$ falls within the observed wavelength ranges masked in Section \ref{subsect:fitting_final}, and is therefore strongly contaminated by sky lines, such that no measurement could be made. For these objects we apply the same fitting methodology to the posterior prediction for this spectral region from our fitted \bagpipes\ model.

Our sample is shown on the EW$_{\mathrm{H}\delta}$ vs D\textsubscript{n}4000 plane in the left panel of Fig. \ref{fig:d4000_hd}. Our sub-samples are significantly offset, with our green valley objects having stronger H$\delta$ absorption and a weaker 4000\AA\ break. There is a clear transition at 1.3~<~D\textsubscript{n}4000~<~1.4 and EW$_{\mathrm{H}\delta}\sim4$\AA, with almost all of our quiescent sub-sample at higher D\textsubscript{n}4000 and lower EW$_{\mathrm{H}\delta}$. Four significant outliers are visible: three quiescent objects with low D\textsubscript{n}4000 and strong H$\delta$ absorption, and one green-valley object with a strong 4000\AA\ break (D\textsubscript{n}4000 $\sim$ 1.55). These objects will be discussed in Section \ref{subsect:results_psbs}.

In the right panel of Fig. \ref{fig:d4000_hd} our quiescent sub-sample is shown coloured by our inferred mass-weighted ages (see Section \ref{subsect:results_age}). A trend in age with D\textsubscript{n}4000 is visible as expected, with our oldest objects having D\textsubscript{n}4000 $\simeq$ 1.7. Contours are plotted showing the distributions of quiescent galaxies at lower observed redshifts. Both samples are taken from \citet[priv. comm.]{Wu2018a} with the additional imposition of our nSFR < 0.1 criterion. The SDSS sample, shown in gray, is at $0.04~<~z~<~0.14$; the Lega-C sample, shown in blue, is at $0.6~<~z~<~1.0$. A similar evolution  of $\sim0.2$ in the average D\textsubscript{n}4000 value can be seen from $z\sim0.1$ to $z\sim0.8$ and from $z\sim0.8$ to our sample at an average redshift of $z\sim1.15$. The cosmic time interval between SDSS and Lega-C is approximately four times that between Lega-C and VANDELS, demonstrating that the distribution of quiescent galaxies moves towards lower D\textsubscript{n}4000 at an accelerating pace with increasing lookback time as expected.

\begin{figure*}
	\includegraphics[width=\textwidth]{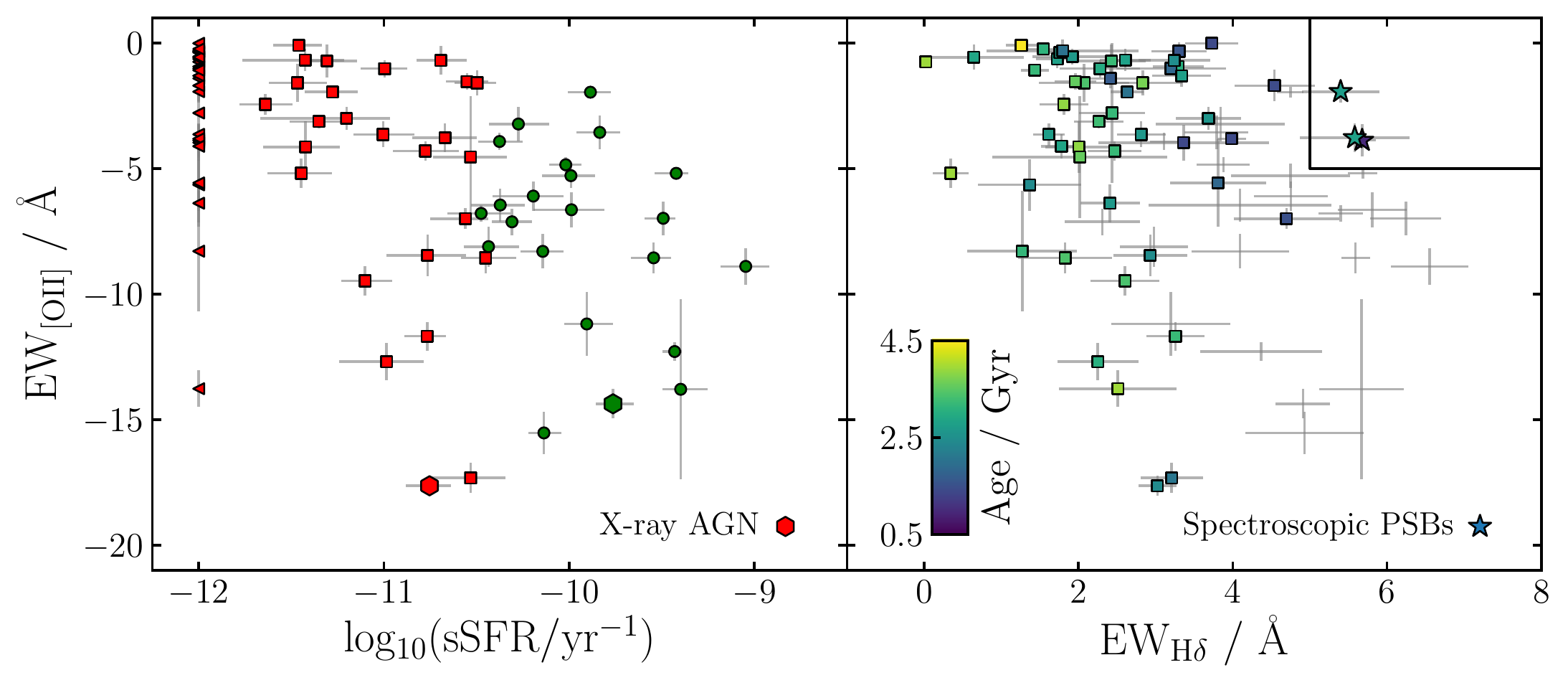}
    \caption{The distribution of [O\,\textsc{ii}] emission in our sample. The left panel shows EW$_\mathrm{[O\,\textsc{ii}]}$ vs sSFR: symbols and colours are as defined in Fig. \ref{fig:d4000_hd}, except for galaxies with log$_{10}$(sSFR/yr$^{-1})~<~-12$, which are shown as triangles, indicating an upper limit. Additionally, objects detected in the \textit{Chandra} seven megasecond catalogue of \protect \cite{Luo2017} are shown as hegaxons (see Section \ref{subsect:results_agn}). The right panel shows the EW$_\mathrm{H\delta}$ vs EW$_\mathrm{[O\,\textsc{ii}]}$ parameter space typically used to select post-starburst galaxies. Quiescent galaxies are coloured by mass-weighted age, green-valley galaxies are shown with gray errorbars. Commonly used spectroscopic PSB selection criteria are shown with solid lines (e.g. \citealt{Tran2003}; \citealt{Maltby2016}). Galaxies which meet these criteria are shown as stars (see Section \ref{subsect:results_psbs}). }\label{fig:oii}
 \end{figure*}

\subsection{Trends with rest-frame UVJ colours}\label{subsect:results_uvj}

In this section we consider the positions of our galaxies on the UVJ diagram, and physical parameter trends with UVJ colours. Our sample is shown on the UVJ diagram in Fig. \ref{fig:uvj}. 

\subsubsection{Trends with star-formation rate}\label{subsubsect:results_uvj_ssfr}

The top-left panel of Fig. \ref{fig:uvj} shows a direct comparison of our nSFR-based selection to both sets of UVJ selection criteria discussed in Section \ref{subsect:results_classification}. Good agreement can be seen between the dashed criterion of $U - V$ > 0.88($V - J$) + 0.69 and the nSFR-based selection introduced by \cite{Carnall2018}. None of our quiescent sub-sample fall below the dashed line, whereas seven green valley objects fall above. All seven are very close to our nSFR threshold, with the majority having 16th posterior percentiles at nSFR < 0.1.

The top-right panel of Fig. \ref{fig:uvj} shows our sample coloured by nSFR. At an observed redshift of $z\sim1$, a nSFR value of 0.1 is equivalent to log$_{10}$(sSFR/yr$^{-1})~\simeq~-10.5$. We observe a clear trend in nSFR (and hence sSFR) perpendicular to the red sequence, in agreement with recent results (e.g. \citealt{Fang2018}; \citealt{Carnall2018}). It is worth noting that the solid box representing the more permissive UVJ selection criteria includes objects which are forming stars at up to $\sim50{-}100$ per cent of their historical average SFRs, meaning that these criteria alone should not be used to select high-redshift quiescent galaxies.

We therefore argue that the nSFR < 0.1 criterion is the most robust method for selecting quiescent galaxy samples. This method is truly redshift-independent, as it selects galaxies which are forming stars below a fixed fraction of their historical average SFRs. Both a fixed UVJ selection of $U - V$ > 0.88($V - J$) + 0.69 and sSFR selection proportional to $t^{-1}$ produce results similar to nSFR < 0.1. By contrast, we argue that both the original redshift-dependent UVJ selection criteria, and selection using a fixed sSFR threshold are less appropriate, as they include galaxies which are proportionally more highly star-forming  at higher redshifts.

\subsubsection{Trends with dust attenuation}\label{subsubsect:results_uvj_dust}

Our sample is shown coloured by $A_\mathrm{V}$ in the bottom-left panel of Fig. \ref{fig:uvj}. It can be seen that we find a strong trend in dust attenuation across our UVJ box, which follows the trend we observe in sSFR, \textit{perpendicular} to the dust reddening vector. These results are in accordance with those of \cite{Fang2018}, who observe a drop in the dust content of their star-forming galaxies as they move closer to the red sequence. We find that the majority of objects within the solid UVJ selection box which have $A_\mathrm{V} > 1$ are those we identify as in the green valley, rather than truly quiescent. This confirms that, within the UVJ box, dust attenuation is more strongly related to sSFR than position along the dust-reddening vector. Our quiescent sub-sample is typically found to be less dusty, however there is still a noticeable trend with distance from the edge of the UVJ selection box, as was found by \cite{Belli2018}.

\subsubsection{Trends with stellar age}\label{subsubsect:results_uvj_age}

The mass-weighted ages of our quiescent sub-sample (see Section \ref{subsect:results_age}) are shown on the UVJ diagram in the bottom-right panel of Fig. \ref{fig:uvj}. Also plotted are ages predicted by the relationship derived by \cite{Belli2018}. The lines which run perpendicular to the dashed UVJ selection are lines of constant age, which are shown from $0.5{-}4.5$ Gyr in intervals of 0.5 Gyr. The ages we derive for our objects are in good agreement with \cite{Belli2018}, despite several methodological differences (median vs mean stellar age, different SFH models), demonstrating that the ages of quiescent galaxies are less model-dependent than those of star-forming objects.

It is remarkable that such a clear trend in stellar age can exist parallel to the dust reddening vector on the UVJ diagram. Galaxies along the top-left edge of our distribution can be seen to follow a pure age sequence, with no evolution in dust attenuation. A population of totally quenched objects with younger stellar populations which still retained significantly more dust would disrupt this, as well as the age trend found by \cite{Belli2018}.  This implies that quenching galaxies must lose most of their dust before their sSFRs drop to the extent where they can join the red sequence. These ideas will be explored further in Section \ref{subsect:discussion_sequence}.

\subsection{[O\,\sc{ii}\bf{] emission properties}}\label{subsect:results_oii}

As described in Section \ref{subsect:fitting_final}, the [O\,\textsc{ii}] 3727\AA\ emission line was masked during our spectral fitting analysis due to uncertainties as to the excitation mechanism in quiescent galaxies. We now consider the distribution of [O\,\textsc{ii}] emission in our sample. We first measure the rest-frame equivalent width of the line, EW$_\mathrm{[O\,\textsc{ii}]}$, using the same method as was applied to the H$\delta$ feature in Section \ref{subsect:results_d4000_hd}. 

A comparison of our measured [O\,\textsc{ii}] equivalent widths with our inferred sSFRs is shown in the left panel of Fig. \ref{fig:oii}. Our green-valley galaxies typically exhibit stronger [O\,\textsc{ii}] emission, with 77 per cent having EW$_\mathrm{[O\,\textsc{ii}]}~<~-5$\AA. By contrast, only 26 per cent of our quiescent sub-sample has [O\,\textsc{ii}] emission stronger than this threshold. This is despite the dustier nature of our green valley sub-sample (see Section \ref{subsubsect:results_uvj_dust}). Whilst our green-valley sub-sample clearly exhibits stronger [O\,\textsc{ii}] than our quiescent sub-sample, it can be seen that at fixed EW$_\mathrm{[O\,\textsc{ii}]}$ our inferred sSFRs span a range as large as $\sim2$ dex. This is in agreement with previous results which identify [O\,\textsc{ii}] as a poor predictor of ongoing star-formation in quiescent galaxies (e.g. \citealt{Lemaux2010}).

Without rest-frame optical spectroscopy it is challenging to constrain possible AGN contributions to our [O\,\textsc{ii}] fluxes. However, the majority of local quiescent galaxies with detectable [O\,\textsc{ii}] emission have been shown to exhibit high [O\,\textsc{ii}]/H$\alpha$ ratios (e.g. \citealt{Yan2006}) which are inconsistent with excitation by ongoing star-formation \citep{Kewley2004}. We therefore identify galaxies in our sample with strong [O\,\textsc{ii}] emission as likely hosts of low-level AGN activity, in particular the quiescent galaxies for which we find EW$_\mathrm{[O\,\textsc{ii}]}~<~-10$\AA. We will further consider the possibility of AGN activity in our galaxies in Section \ref{subsect:results_agn}.

Ionization by hot low-mass stars has also been postulated as an explanation for line emission in quiescent galaxies (e.g. \citealt{Singh2013}). Recently, \cite{Herpich2018} reported a marginal difference in the stellar ages of local quiescent galaxies with and without visible emission lines. In principle, this measurement should be easier to make at $z\sim1$, as the stellar populations of quiescent galaxies are considerably younger. Our quiescent sub-sample is shown coloured by stellar age in the right panel of Fig. \ref{fig:oii}. No clear correlation between [O\,\textsc{ii}] emission and stellar age is visible, however our sample is considerably smaller than those available in the local Universe, which may preclude the detection of this subtle effect. Future large high-redshift spectroscopic surveys will be a valuable tool for addressing this issue.

\subsection{Post-starburst and rejuvenated galaxies}\label{subsect:results_psbs}

Post-starburst galaxies (PSBs) are widely identified as one of two major transitional states between the star-forming population and red sequence (see Section \ref{sect:intro}). A variety of methods have been used to identify samples of PSBs, ranging from spectroscopic selection based on strong H$\delta$ or H$\beta$ absorption and a lack of emission lines (e.g. \citealt{Tran2003}), to principal component analyses (e.g. \citealt{Wild2007, Wild2014}), to selection by rest-frame UVJ magnitudes (e.g. \citealt{Belli2018}). Comparisons between different methods have found significant overlap (e.g. \citealt{Maltby2016}), however a fully self-consistent set of criteria which returns objects with the desired properties is still to be agreed upon.

A detailed discussion of the physical properties of VANDELS PSBs will be presented by Wild et al. in prep. In this section we briefly discuss the distribution of PSBs within our sample, and the degree of consistency between different selection methods. The right panel of Fig. \ref{fig:oii} shows the EW$_\mathrm{[O\,\textsc{ii}]}$ vs EW$_\mathrm{H\delta}$ parameter space often used to spectroscopically select PSBs (e.g. \citealt{Tran2003}; \citealt[submitted]{Maltby2016}). We identify three spectroscopic PSBs, all of which are members of our quiescent sub-sample. These objects are marked with stars in Figs \ref{fig:ages}, \ref{fig:d4000_hd}, \ref{fig:uvj} and \ref{fig:oii}.

Our spectroscopically identified PSBs are shown on the UVJ diagram in the top-right panel of Fig. \ref{fig:uvj}. It can be seen that our PSBs occupy the region predicted by \cite{Wild2014}, towards the bottom-left of the UVJ selection box. The larger of the two blue boxes is the PSB selection used by \cite{Belli2018}, which assumes that PSB features are visible for quiescent galaxies with median stellar ages of $300{-}800$ Myr. Both of the objects we find within this box are part of our green-valley sub-sample and, whilst both are close to our PSB selection box in Fig. \ref{fig:oii}, neither fulfils our PSB selection criteria. The object close to the top of the box has strong H$\delta$ absorption, however it has EW$_\mathrm{[O\,\textsc{ii}]} = -6.8\pm0.3$\AA. By contrast, the object at the bottom of the box fulfils our [O\,\textsc{ii}] criterion but has EW$_\mathrm{H\delta}\sim3$\AA. 

The objects we identify spectroscopically as PSBs occupy a region slightly further up the red sequence, within the region for which \cite{Belli2018} predict ages of $800{-}1200$ Myr. This region is highlighted with a smaller blue box above the one used by \cite{Belli2018}. Our results suggest this is the region in which spectroscopic features usually associated with PSBs are strongest. The fourth object within our extended box, as well as the closest object above this box, have [O\,\textsc{ii}] consistent with our selection criteria but slightly weaker H$\delta$ absorption. These results are in good agreement with \cite{Maltby2016}, who find that $\sim50$ per cent of galaxies identified by the photometric selection proposed by \cite{Wild2014}, upon which the \cite{Belli2018} UVJ selection is based, exhibit [O\,\textsc{ii}] and H$\delta$ equivalent widths consistent with the spectroscopic criteria shown in Fig. \ref{fig:oii}.

Our findings suggest that the contribution of the PSB quenching channel to the growth of the red sequence is towards the upper end of the range found by \cite{Belli2018}. However, as will be discussed in Section \ref{subsect:discussion_sequence}, the timescale over which PSB features are visible is not necessarily determined by the time taken for a galaxy to traverse the blue box on Fig. \ref{fig:uvj} by passive evolution of its stellar population.

We finally note, based upon Fig. \ref{fig:d4000_hd}, the presence of one galaxy in our green valley sub-sample which displays D\textsubscript{n}4000 > 1.5, but which has a blue continuum below this wavelength, and strong [O\,\textsc{ii}] emission. For this object we find a stellar mass of log$_{10}(M_*$/M$_\odot$) = $11.25\pm0.15$ and a mass-weighted age of $3.6\pm0.5$ Gyr, amongst the oldest and most massive in our sample. With currently available data it is challenging to discriminate between rejuvenated star-formation and AGN activity (though we find no evidence of AGN activity in either X-ray or radio datasets in Section \ref{subsect:results_agn}). However, in either case, objects such as this are clearly of significant interest for assisting our understanding of continuing mass assembly in the oldest galaxies (e.g. \citealt{Belli2017}; \citealt{Nelson2018}). Assuming no AGN contribution we infer a SFR of $8.9^{+4.9}_{-3.2}$~M$_\odot$ yr$^{-1}$, meaning the stellar mass of this object would increase by $\sim2{-}5$ per cent over 100 Myr.

\begin{figure*}
	\includegraphics[width=\textwidth]{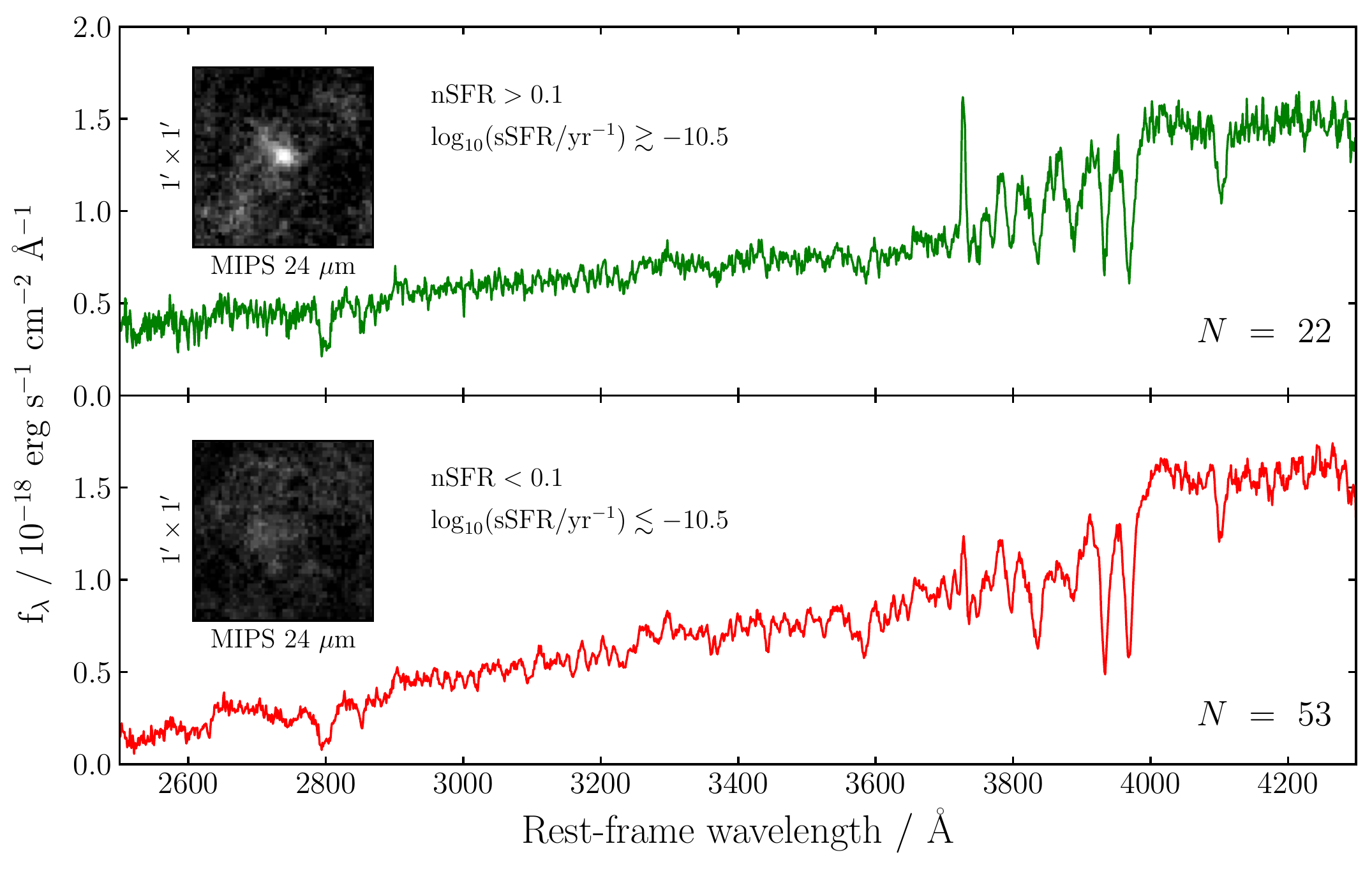}
    \caption{Stacked spectra for our green valley and quiescent sub-samples. Key differences include more flux below 3000\AA\ and stronger [O\,\textsc{ii}] emission for the green-valley galaxies, and a transition from Balmer to 4000\AA\ break. Stacked 24 $\mathrm{\mu}$m images are also shown, which include all 57 objects with MIPS coverage. Our green valley stack shows a clear detection, consistent with greater ongoing star-formation.}\label{fig:stacks}
\end{figure*}

\begin{figure*}
	\includegraphics[width=\textwidth]{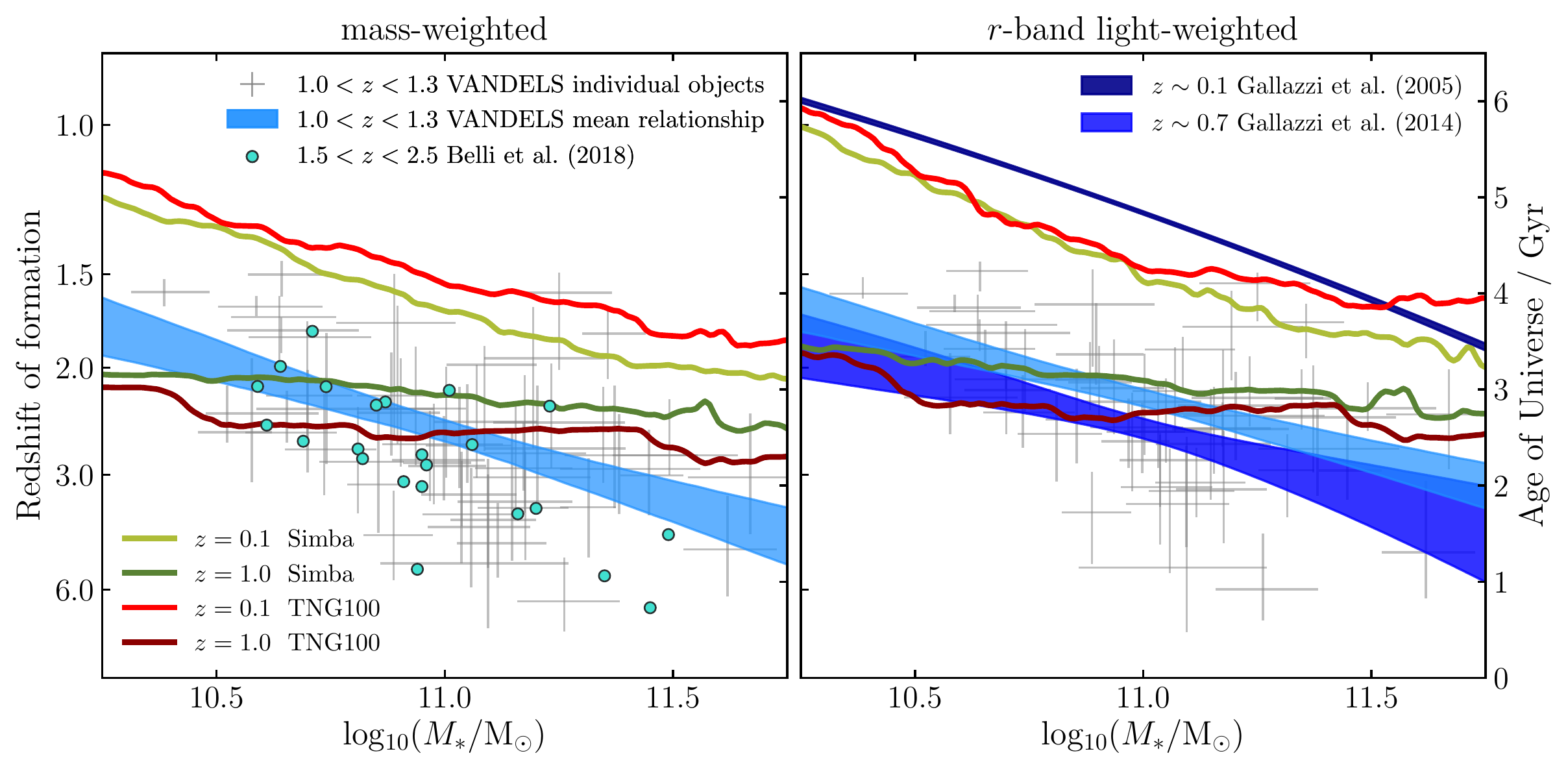}
    \caption{A comparison of quiescent galaxy formation redshifts from spectroscopic studies and simulations at a range of observed redshifts. Formation redshifts in the left panel are mass-weighted, those in the right panel are $r$-band light-weighted. Results at $z=0.1$ are measured within $3^{\prime\prime}-$diameter circular apertures. Observational results at  higher redshifts were obtained using $1^{\prime\prime}$ slits. Results from simulations at $z=1$ were extracted using $1^{\prime\prime}$ square apertures. Systematic offsets exist in the vertical positions of the different observational relationships (e.g. due to the use of different SFH models), however the gradients can be seen to be in good agreement.}\label{fig:tform_spec}
\end{figure*}

\subsection{Evidence of AGN activity}\label{subsect:results_agn}

As discussed in Section \ref{subsect:results_oii}, it is challenging to determine whether or not galaxies in our sample host an AGN, given only our UV-NIR photometry and rest-frame UV spectroscopy. Whilst line ratios from rest-frame optical spectroscopy would be the ideal tool for studying this, we here consider ancillary datasets in the X-ray, mid-infrared and radio to attempt to constrain any AGN contributions.

We first consider X-ray data from the \textit{Chandra} seven megasecond source catalogue \citep{Luo2017} and Subaru/\textit{XMM-Newton} deep survey \citep{Ueda2008}. We find two matches within our sample, both from the \textit{Chandra} seven megasecond source catalogue. These objects are shown as hexagons in the left panel of Fig. \ref{fig:oii}, and can be seen to have two of the four strongest [O\,\textsc{ii}] lines in our sample (the second highest is in the UDS). This suggests that strong [O\,\textsc{ii}] emission is a good predictor of X-ray AGN activity in quiescent galaxies.

We then consider radio data at 1.4GHz from the Very Large Array, which is available for both of our fields (\citealt{Simpson2006}; \citealt{Bonzini2013}). We find that three of our galaxies are detected, none of which are in common with our X-ray detected objects. All three are massive, with log$_{10}(M_*$/M$_\odot$) > $11$, and all three sit close to our nSFR threshold, with log$_{10}$(sSFR/yr$^{-1})~\simeq~-10.5$. There is no indication from their rest-frame UV continua or [O\,\textsc{ii}] lines that these galaxies host AGN, as has typically been found for radio galaxies at high redshift (e.g. \citealt{Dunlop1996}).

We finally consider publicly available $Spitzer$ MIPS 24$\mathrm{\mu}$m imaging, which is available for 57 of our galaxies (\citealt{Dickinson2003}; \citealt{Dunlop2007}). For each of these objects we extract fluxes within 14$^{\prime\prime}-$diameter apertures and manually inspect the images, flagging objects as isolated (33 objects) or potentially confused (24 objects). Of our isolated sources, 21 are members of our quiescent sub-sample, and 12 are members of our green valley sub-sample.

Only one of the two objects identified above as X-ray sources has MIPS 24$\mathrm{\mu}$m coverage: the galaxy in our green valley sub-sample. This object has the strongest isolated detection in our sample by approximately a factor of two. Of our 21 isolated quiescent galaxies, only one has a detection at >3$\sigma$, whereas 7 out of 12 isolated green valley objects are detected above this threshold. For our eight objects with robust, isolated detections we use the calibration of \cite{Kennicutt2012} to convert our aperture-corrected fluxes into SFRs. We find that the X-ray detected source is the only object with significantly more 24$\mathrm{\mu}$m-flux than expected, based on the SFRs we derive from our spectral fitting analysis.

The quiescent object which is detected at 24$\mathrm{\mu}$m is one of the three identified as radio AGN above. This object is the most massive in our sample, with log$_{10}(M_*$/M$_\odot$) = $11.66\pm0.14$, and one of the oldest, with a mass-weighted age of 3.5$\pm$0.6 Gyr. However, we find no evidence of an AGN contribution to its 24$\mathrm{\mu}$m flux: our inferred SFR is $13.2^{+5.6}_{-3.8}$~M$_\odot$~yr$^{-1}$, whereas its 24$\mathrm{\mu}$m-predicted SFR is $14.3\pm2.5$~M$_\odot$~yr$^{-1}$. This object is similar to the rejuvenated object discussed in Section \ref{subsect:results_psbs}, and seems consistent with the scenario proposed by \cite{Best2014}, in which cooling of gas in the hot halo begins to provide fuel for jet-mode AGN activity and rejuvenated star-formation $\gtrsim2$ Gyr after quenching has occurred. 

The two 1.4GHz catalogues we consider jointly provide coverage down to 100$\mu$Jy sensitivity for all our objects. This corresponds to a rest-frame 1.4GHz luminosity of $10^{23.6}$ W\,Hz$^{-1}$ at $z=1$, or $10^{23.9}$ W\,Hz$^{-1}$ at $z=1.3$. We find that 2 of our 4 objects with log$_{10}(M_*$/M$_\odot$) > $11.5$ are detected above this level, whereas only 1 of our 39 objects  with 11.0 < log$_{10}(M_*$/M$_\odot$) < $11.5$ is detected. This is broadly consistent with the local relationship between radio-loud AGN fraction and stellar mass (e.g. \citealt{Best2005}; \citealt{Sabater2019}), in agreement with other studies which find little redshift evolution (e.g. \citealt{Tasse2008}; \citealt{Simpson2013}).

\subsection{Stacking analyses}\label{subsect:results_stacking}

We finally perform a stacking analysis to demonstrate the average spectral properties of the galaxies in our two sub-samples. We de-redshift our spectra and normalise over the rest-frame wavelength range from $3200{-}3600$\AA, which is in the centre of our observed spectral range and contains no strong features. We then resample our spectra to a common wavelength grid using \textsc{Spectres} \citep{Carnall2017}. Median stacked green valley and quiescent spectra are shown in the top and bottom panels of Fig. \ref{fig:stacks} respectively. The stacks have total exposure times of 745 and 1971 hours respectively.

A clear transition from Balmer to 4000\AA\ break can be seen between the two spectra, demonstrating the older stellar populations of our quiescent galaxies. As discussed in Section \ref{subsect:results_oii}, the median [O\,\textsc{ii}] flux is significantly higher in our green valley sub-sample. The continuum below 4000\AA\ can also be seen to be bluer, with considerably more flux at $<3000$\AA, indicating higher levels of ongoing star-formation.

Median stacked 24$\mathrm{\mu}$m images are also shown in Fig. \ref{fig:stacks}, using all 57 objects for which MIPS coverage is available (see Section \ref{subsect:results_agn}). Our green valley stack shows a clear detection, with no corresponding detection in our quiescent stack.

\section{Discussion}\label{sect:discussion}

In this section we further discuss our results. In Section \ref{subsect:discussion_downsizing} we compare our stellar mass vs age relationship to results from the literature and recent simulations. In Section \ref{subsect:discussion_sequence}, we consider the connections between our quiescent, green-valley and post-starburst galaxies. In Section \ref{subsect:discussion_sfhs} we discuss the shapes of the SFHs we infer for our quiescent sub-sample.

\subsection{The stellar mass vs stellar age relationship}\label{subsect:discussion_downsizing}

As discussed in Section \ref{sect:intro}, the epoch of formation as a function of galaxy stellar mass is a key observable property which is strongly constraining on AGN-feedback models. In this section we compare our results to other observational studies, as well as making comparisons with simulations. 

\subsubsection{Stellar mass vs age from spectroscopic studies}\label{subsubsect:discussion_downsizing_spectroscopy}

The stellar mass vs stellar age relationship we derive for our quiescent sub-sample is compared to results from the literature in Fig. \ref{fig:tform_spec}. In the left panel we show the mean relationship we derive between stellar mass and mass-weighted formation time for our quiescent sub-sample in Section \ref{subsect:results_age}, along with the positions of our individual quiescent galaxies.

In order to facilitate comparisons with earlier work, we also calculate the same mean relationship using $r$-band light-weighted formation times, $t_{r\mathrm{{\text -}band}}$ (e.g. \citealt{Gallazzi2005}). For this quantity we obtain a mean relationship of
\begin{equation}\label{eqn:trband}
\bigg(\dfrac{t_{r\mathrm{{\text -}band}}}{\mathrm{Gyr}}\bigg)\ =\ 2.91^{+0.08}_{-0.09}\ -\ 1.24^{+0.27}_{-0.30}\ \mathrm{log}_{10}\bigg(\dfrac{M_*}{10^{11}\mathrm{M_\odot}}\bigg)
\end{equation}

\noindent with an intrinsic scatter of $0.51^{+0.08}_{-0.07}$ Gyr. The slope of this relationship is slightly shallower than that which we obtain for $t_\mathrm{form}$ (Equation \ref{eqn:tform}), and is offset towards later formation times by $\sim350$ Myr at $\mathrm{log}_{10}(M_*/\mathrm{M_\odot}) = 11$.

In the left panel of Fig. \ref{fig:tform_spec}, we also show the 23 galaxies studied by \cite{Belli2018} at observed redshifts of $1.5 < z < 2.5$. We have shown in Section \ref{subsubsect:results_uvj_age} that our inferred ages agree well with the predictions of \cite{Belli2018}, meaning that no significant systematic offsets should exist between the two samples. A similar range of formation redshifts can be seen for both samples, with the \cite{Belli2018} objects appearing slightly offset towards earlier formation redshifts on average. The slope of the relationship can be seen to be similar in both studies.

In the right panel our $r$-band light-weighted formation times are compared to those found by \cite{Gallazzi2005, Gallazzi2014} for quiescent galaxies at $z\sim0.1$ and $z\sim0.7$ respectively. Again, good agreement can be seen between the slope we infer for our relationship and those found by \cite{Gallazzi2005, Gallazzi2014}. We thus conclude that, at log$_{10}(M_*$/M$_\odot$) > 10.3, an evolution of $\sim1.5$ Gyr in formation time per decade in stellar mass is a robust result which remains consistent from the local Universe to at least $z\sim2$.

However, the relative vertical positions of our relationship and the \cite{Gallazzi2014} relationship at $z\sim0.7$ do not follow the expected trend of earlier formation with increasing observed redshift at fixed stellar mass. This is likely to be due to the significant methodological differences between the two studies. \cite{Gallazzi2014} fit indices derived from rest-frame optical spectra and use an exponentially declining SFH model, whereas we apply a full-spectral-fitting approach to rest-frame UV spectra using a double-power-law SFH model. The use of different SFH models and priors is known to affect derived ages, which could plausibly give rise to this effect (e.g. \citealt{carnall2019}; \citealt{Leja2019a, Leja2019b}). It is additionally likely that the optical spectra of quiescent galaxies are dominated by an older population than dominates in the UV, meaning that fitting different spectral regions could result in different SFHs. 

\begin{figure}
	\includegraphics[width=\columnwidth]{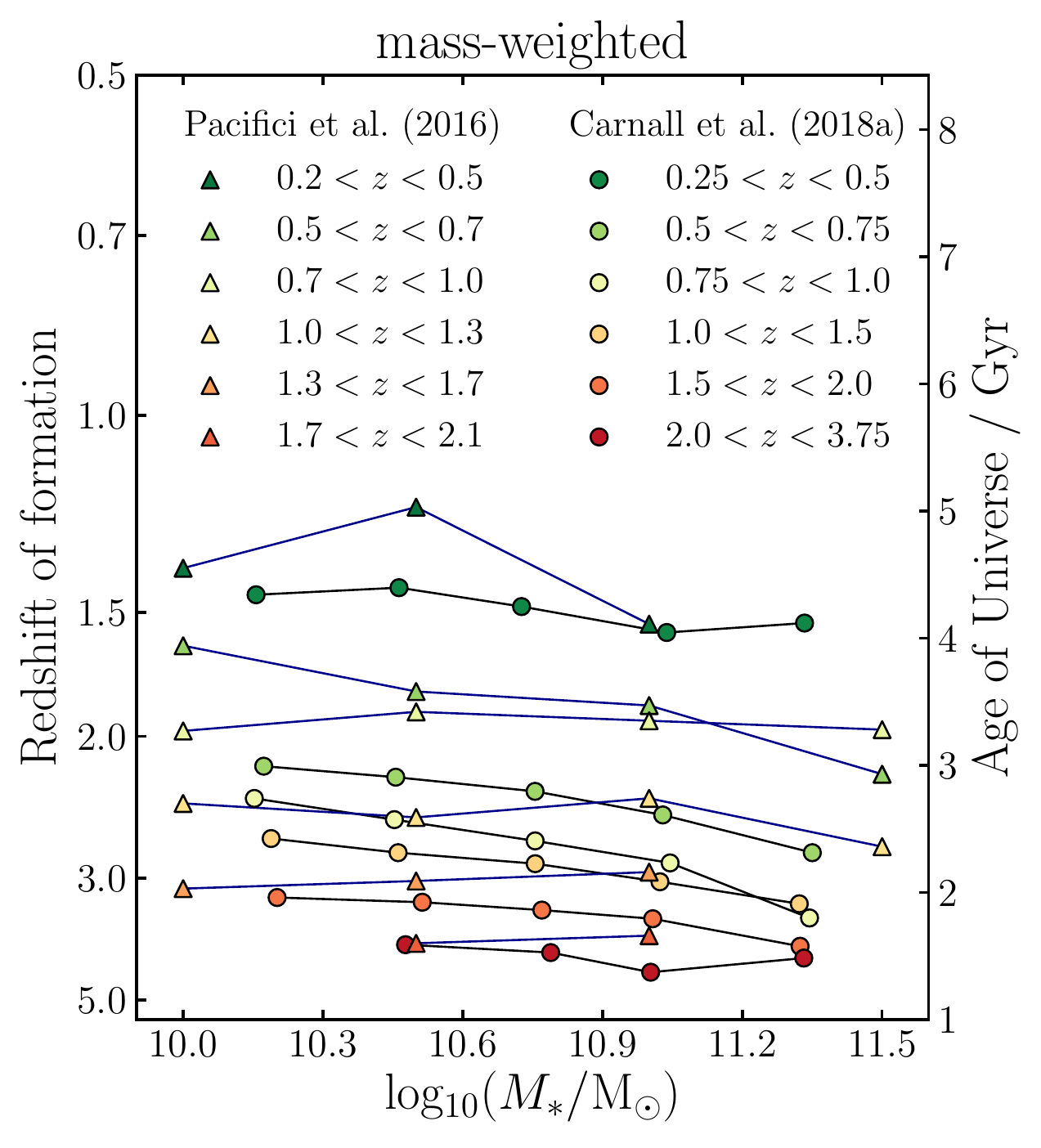}
    \caption{A comparison of formation redshifts for massive quiescent galaxies from two recent photometric studies. Whilst these two sets of results are in good agreement, considerably weaker stellar mass vs stellar age trends are recovered when compared to the spectroscopic studies shown in Fig. \ref{fig:tform_spec}.}\label{fig:tform_phot}
\end{figure}

\subsubsection{Stellar mass vs age in cosmological simulations}\label{subsubsect:discussion_downsizing_simulations}

Given our conclusion in Section \ref{subsubsect:discussion_downsizing_spectroscopy} that the steep stellar-mass vs stellar age relationship observed for quiescent galaxies at low redshift is already in place by $z\sim2$, it is interesting to consider whether this trend is reproduced by modern cosmological simulations. Historically, this relationship has been challenging to match, even in the local Universe, for both quiescent and star-forming galaxies (e.g. \citealt{Somerville2008}; \citealt{Trager2009}).

We consider predictions from the 100 $h^{-1}$ Mpc box length runs of both \textsc{Simba} \citep{Dave2019} and \textsc{IllustrisTNG} (e.g. \citealt{Nelson2019}), using snapshots at $z=0.1$ and $1.0$. In order to match these predictions as closely as possible with the observational studies discussed in Section \ref{subsubsect:discussion_downsizing_spectroscopy}, we apply apertures to the simulated galaxies as follows. For the $z=0.1$ snapshots, we apply $3^{\prime\prime}-$diameter circular apertures to each galaxy, for consistency with the SDSS observations used by \cite{Gallazzi2005}. For the $z=1$ snapshots, we apply $1^{\prime\prime}$ square apertures, to match the $1^{\prime\prime}$ slits used by \cite{Gallazzi2014}, \cite{Belli2018} and this work (assuming a $\sim1^{\prime\prime}$ region along the direction of the slit is extracted). We select all galaxies within these snapshots which meet our nSFR < 0.1 criterion (see Section \ref{subsect:results_age}), then calculate mass-weighted and $r$-band light-weighted formation times for the simulated galaxies. We use \bagpipes\ to predict the $r$-band flux from each star particle.

In Fig. \ref{fig:tform_spec} we show the average mass-weighted and $r$-band light-weighted formation times as a function of stellar mass for each snapshot. At each point along the horizontal axis, the median formation time of galaxies within a 0.25 dex mass range centred on that point is shown. The two simulations can be seen to be in reasonably good agreement, with discrepancies confined to levels of $\lesssim250$ Myr. The $z=0.1$ relationships predicted by these simulations have slopes consistent with the $\sim1.5$ Gyr per decade in mass evolution found by the observational studies shown. Additionally, the normalisations of the $z=0.1$ relationships in both simulations are in good agreement with the results of \cite{Gallazzi2005}, as found for the general $z<0.2$ galaxy population in \textsc{IllustrisTNG} by \cite{Nelson2018}.

However, at $z=1$ both simulations predict significantly weaker average stellar mass vs age relationships. Using mass-weighted formation times, \textsc{Simba} predicts an evolution of 0.38 Gyr per decade in stellar mass across the interval shown in Fig. \ref{fig:tform_spec}, whereas \textsc{IllustrisTNG} predicts an evolution of 0.51 Gyr per decade. This suggests that these simulations do not accurately reproduce the detailed physical properties of massive quiescent galaxies at $z>1$. This is particularly interesting in the context of the results of \cite{Schreiber2018}, who find that the precursors to these simulations (\textsc{Mufasa} and \textsc{Illustris}) significantly under-predict the number density of quiescent galaxies at $3<z<4$.

\subsubsection{Stellar mass vs age from photometric studies}\label{subsubsect:discussion_downsizing_photometry}

Several recent studies also attempt to probe the stellar mass vs age relationship using photometric data. This has the advantage of providing better statistics, as larger samples are available. However, as discussed in Section \ref{sect:intro}, the age-metallicity-dust degeneracy leads to larger uncertainties on individual age measurements. In this section we compare two recent photometric studies to the spectroscopic studies discussed in Section \ref{subsubsect:discussion_downsizing_spectroscopy}. 

\cite{Pacifici2016} consider a sample of 845 objects with multi-band photometry from CANDELS, whereas \cite{Carnall2018} consider a sample of 9289 galaxies from UltraVISTA \citep{McCracken2012}. The results of these two studies are summarised in Fig. \ref{fig:tform_phot}. We derive $t_\mathrm{form}$ for the stacked SFHs of \cite{Pacifici2016} using the best-fitting double-power-law parameters reported in their table A1. Despite their very different methodologies, these studies can be seen to produce similar results. However, the slopes reported are considerably shallower than those shown in Fig. \ref{fig:tform_spec}, at $\lesssim0.5$ Gyr per decade in stellar mass.

We attribute this inconsistency to the increased uncertainties on individual object SFHs when considering photometric data. In the presence of large uncertainties, a population of very massive, very old objects will be preferentially scattered towards later formation times.  This is due to the constraint imposed by the age of the Universe. Similarly, a population of younger, less massive quiescent galaxies will be preferentially scattered towards earlier formation times, as the time-evolution of galaxy stellar populations is much more rapid at younger ages, meaning later formation times will be strongly inconsistent with the observed spectrum.

Both of these effects act to flatten the trends observed in Fig. \ref{fig:tform_phot} when compared to the spectroscopic analyses in Fig. \ref{fig:tform_spec}. This finding highlights the importance of upcoming large spectroscopic surveys in providing stronger constraints on stellar ages than are available from photometric data alone (e.g. \citealt{Pacifici2012}; \citealt{Thomas2017}).

\begin{figure}
	\includegraphics[width=\columnwidth]{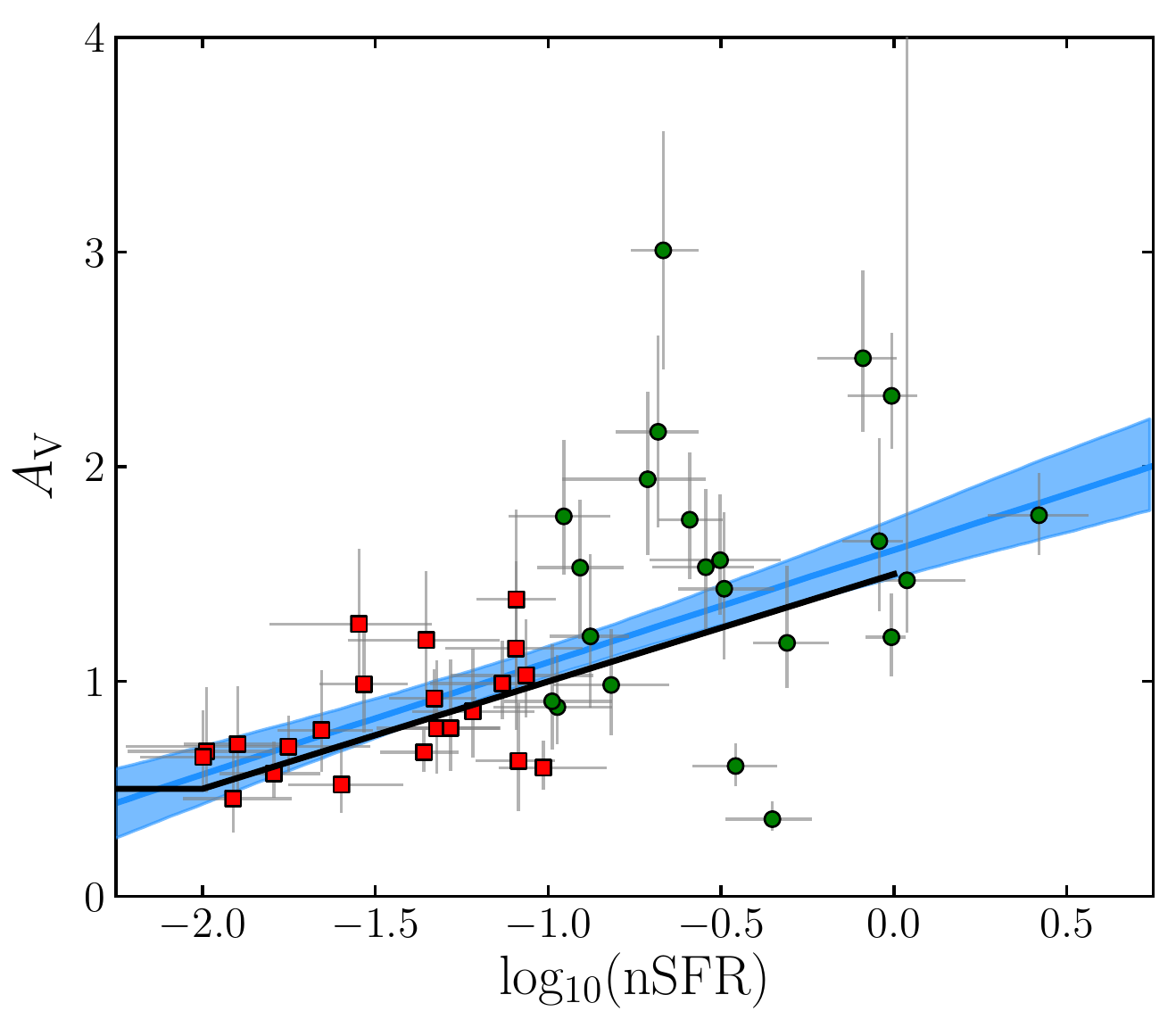}
    \caption{Dust attenuation vs nSFR for galaxies in our sample with detectable star-formation (nSFR > 0.01). The blue line shows the posterior median linear relationship reported in Equation \ref{eqn:dust}, whereas the black line shows the ansatz used in Fig. \ref{fig:uvj_tracks}.}\label{fig:dust_nsfr}
\end{figure}

\subsection{Connecting green-valley, post-starburst and quiescent galaxies}\label{subsect:discussion_sequence}

Much debate exists as to how galaxies evolve away from an initial position within the star-forming population, potentially through green-valley or post-starburst phases, to eventually join the red sequence. Several recent studies have considered this problem in the context of evolutionary tracks across the UVJ diagram (e.g. \citealt{Belli2018}; \citealt{Morishita2018}). One of the most challenging aspects of this approach is modelling the time-evolution of dust attenuation, which has a significant impact on these evolutionary tracks.

\cite{Belli2018} consider the number of galaxies which pass through the larger blue PSB selection box drawn on the top-right panel of Fig. \ref{fig:uvj}, arguing for separate fast and slow quenching mechanisms which do and do not pass through the box respectively. They find that fast quenching plays a more important role at high redshift, in accordance with previous work (e.g. \citealt{Wild2009, Wild2016}; \citealt{Schawinski2014}; \citealt{Pacifici2016}; \citealt{Carnall2018}). A toy model to describe both fast and slow quenching routes is presented in their fig. 12, assuming that $A_\mathrm{V} \propto$ SFR. For their exponentially declining SFH models this means galaxies lose most of their dust early on, and the tracks therefore approach the UVJ selection box from the bottom-left part of the diagram. 

However, for our green-valley sub-sample we find relatively high dust attenuation ($A_\mathrm{V} \sim1{-}2$) even very close to the dashed UVJ boundary in Fig. \ref{fig:uvj}. This implies that these objects evolve differently to the scenarios proposed by \cite{Belli2018}, approaching the UVJ selection box from further upwards and to the right on the UVJ diagram. This is expected, given their high masses and the fact that $A_\mathrm{V}$ is more strongly correlated with stellar mass than SFR in star-forming galaxies (e.g. \citealt{Garn2010}; \citealt{McLure2018a}). In this section we discuss quenching scenarios for green-valley galaxies.

\begin{figure*}
	\includegraphics[width=\textwidth]{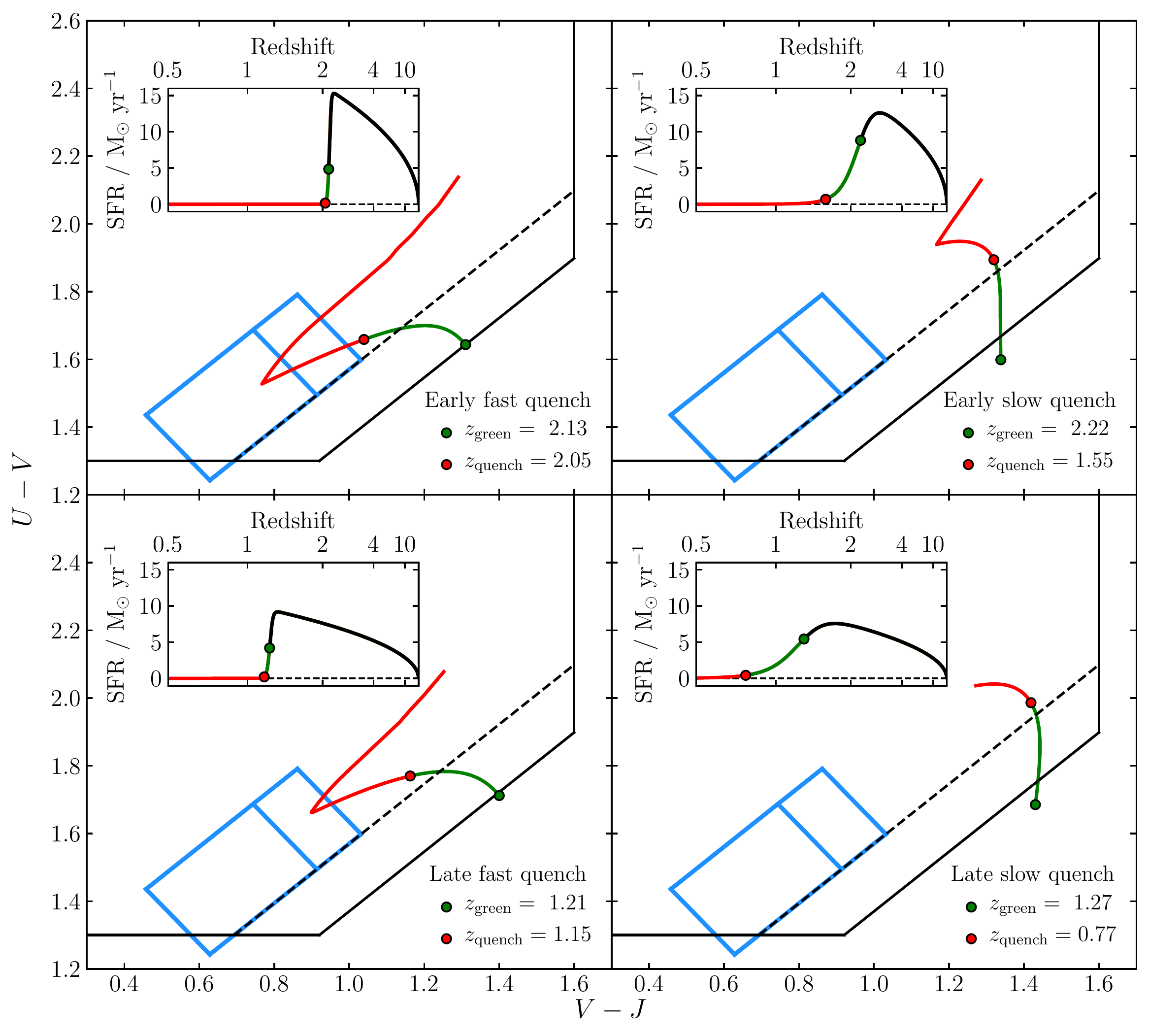}
    \caption{UVJ tracks for four representative galaxy SFHs introduced in Section \ref{subsubsect:discussion_sequence_model}, assuming the relationship between dust attenuation and nSFR shown in black in Fig. \ref{fig:dust_nsfr}. The SFH for each object is shown in the inset panel. The solid and dashed black lines, as well as the blue PSB selection box, are as described in the caption of Fig. \ref{fig:uvj}. The green points show the redshifts at which these objects enter the green valley (nSFR = 1; when we begin tracking them). The red points show the redshifts at which they quench (nSFR = 0.1).}\label{fig:uvj_tracks}
\end{figure*}

\subsubsection{Modelling the evolution of UVJ colours}\label{subsubsect:discussion_sequence_model}

We begin by considering the time-evolution of dust attenuation. As discussed in Section \ref{subsect:results_uvj}, we observe strong trends in both nSFR and $A_\mathrm{V}$ perpendicular to the dashed UVJ selection in Fig. \ref{fig:uvj}, meaning nSFR and $A_\mathrm{V}$ are correlated. This relationship is shown in Fig. \ref{fig:dust_nsfr} for galaxies with detectable levels of star-formation (nSFR > 0.01; see Section \ref{subsect:results_age}). We fit a linear relationship as described in Section \ref{subsect:results_age}, obtaining
\begin{equation}\label{eqn:dust}
A_\mathrm{V} = 0.52^{+0.12}_{-0.11}\ \mathrm{log}_{10}(\mathrm{nSFR}) + 1.61^{+0.15}_{-0.13}
\end{equation}

\noindent with an intrinsic scatter of $0.37^{+0.06}_{-0.05}$ magnitudes. At lower nSFR we find that $A_\mathrm{V}$ reaches a minimum value of $\sim 0.5$ as discussed in Section \ref{subsubsect:results_uvj_dust}. At higher nSFR we do not have galaxies in our sample to test whether this relationship holds. In the discussion which follows we use a simple empirical ansatz for $A_\mathrm{V} $, consistent with Equation \ref{eqn:dust}. We assume that $A_\mathrm{V} = 0.5\ \mathrm{log}_{10}(\mathrm{nSFR}) + 1.5$ for 0.01 < nSFR < 1, then that $A_\mathrm{V}$ remains constant at 0.5 for lower nSFR (shown as a black line in Fig. \ref{fig:dust_nsfr}). For simplicity we assume the \cite{Calzetti2000} dust attenuation law, with attenuation doubled towards stars formed in the last 10 Myr, as in Section \ref{subsect:model_dust}.

Armed with this relationship, we construct four representative double-power-law SFHs (see Equation \ref{eqn:DPL}) to describe a range of quenching scenarios. We consider ``early quenching'', which occurs at $z\sim2$, and ``late quenching'' at $z\sim1$. For each of these two scenarios we consider fast and slow quenching paths, which have timescales of $\sim100$ Myr and $\sim1$ Gyr respectively, in accordance with \cite{Belli2018}. The parameters of the four models are

\begin{itemize}
\item Early fast quench: $\tau=3$ Gyr, $\beta=0.5$, $\alpha=100$
\smallskip
\item Early slow quench: $\tau=3$ Gyr, $\beta=0.5$, $\alpha=10$
\smallskip
\item Late fast quench: $\tau=5$ Gyr, $\beta=0.5$, $\alpha=100$
\smallskip
\item Late slow quench: $\tau=5$ Gyr, $\beta=0.5$, $\alpha=10$.
\end{itemize}

\noindent We model the evolution of the UVJ colours of these four mock galaxies forwards from the point at which their nSFR first falls below 1. this can be thought of as the time at which they enter the green valley, as this is roughly the highest nSFR we find in our green valley sub-sample (see Fig. \ref{fig:dust_nsfr}).

\subsubsection{Relating the green-valley, post-starburst and quiescent populations}\label{subsubsect:discussion_sequence_results}

The tracks the four mock galaxies follow across the UVJ diagram are shown in Fig. \ref{fig:uvj_tracks}. Their SFHs are shown in the inset panels. We highlight with green and red points the redshifts at which nSFR falls below 1 and 0.1 respectively, corresponding to the times at which these galaxies enter the green valley and then quench. It can be seen that each of these objects at first follows a curving path, the shape of which is determined by both quenching speed and the duration of star-formation activity before quenching begins (as well as depending strongly on the assumptions we have made regarding dust attenuation). At the end of this curving track, galaxies begin to passively age along a straight path towards the upper right of the UVJ box. The ``late slow quench" model has not reached this point by $z=0.5$.

Galaxies which quench rapidly can be seen to briefly pass through the blue PSB selection box, both entering and leaving by the top-right edge. This picture is distinct from the PSB evolution channel of \cite{Belli2018}, which enters the PSB box from the lower-left edge. This is a consequence of different assumptions regarding the evolution of dust: we assume that galaxies lose their dust rapidly during quenching, whereas in the \cite{Belli2018} scenario dust attenuation is already low on approach to the green valley. How far our rapidly quenched galaxies enter into the PSB box depends on how extended in time star-formation activity is prior to quenching, as well as quenching speed. At high redshift, when star-formation cannot be very extended, galaxies fall further into the box, and spend more time in the PSB phase. At lower redshifts, the older stellar populations present in galaxies prevent their UVJ colours becoming blue enough to fall as deep into the PSB box.

\begin{figure}
	\includegraphics[width=\columnwidth]{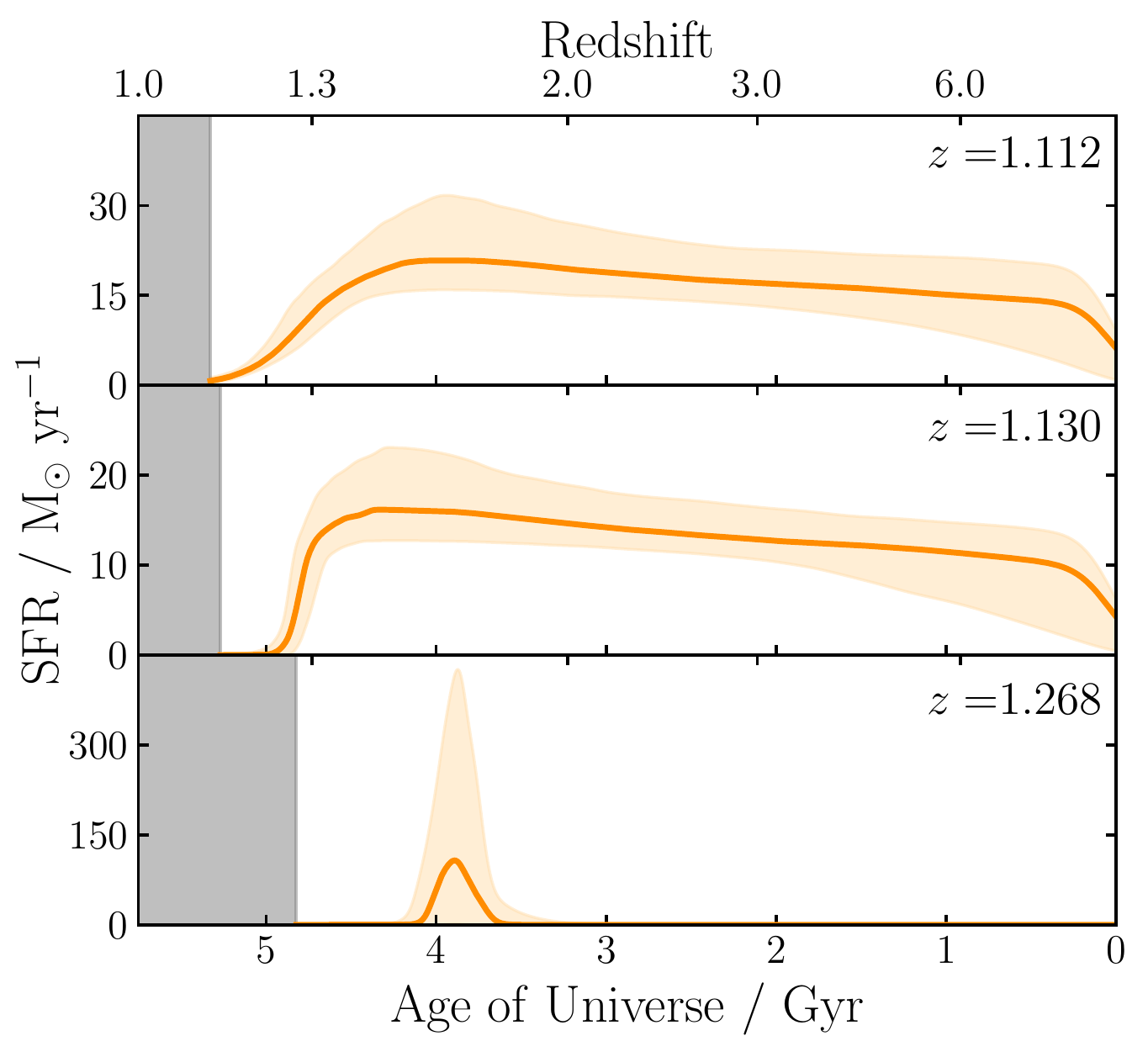}
    \caption{SFHs inferred for the three spectroscopic PSBs identified in Fig. \ref{fig:oii}. Time after the observation of each galaxy is shaded gray. Two have SFHs which are significantly extended before rapidly quenching; the third is a literal post-starburst, formed at $z\sim1.6$. The first two are consistent with the ``late fast quench" scenario shown in Fig. \ref{fig:uvj_tracks}. The third probably evolves similarly to the PSB track shown in fig. 12 of \protect \cite{Belli2018}.}\label{fig:psb_sfhs}
\end{figure}

\begin{figure}
	\includegraphics[width=\columnwidth]{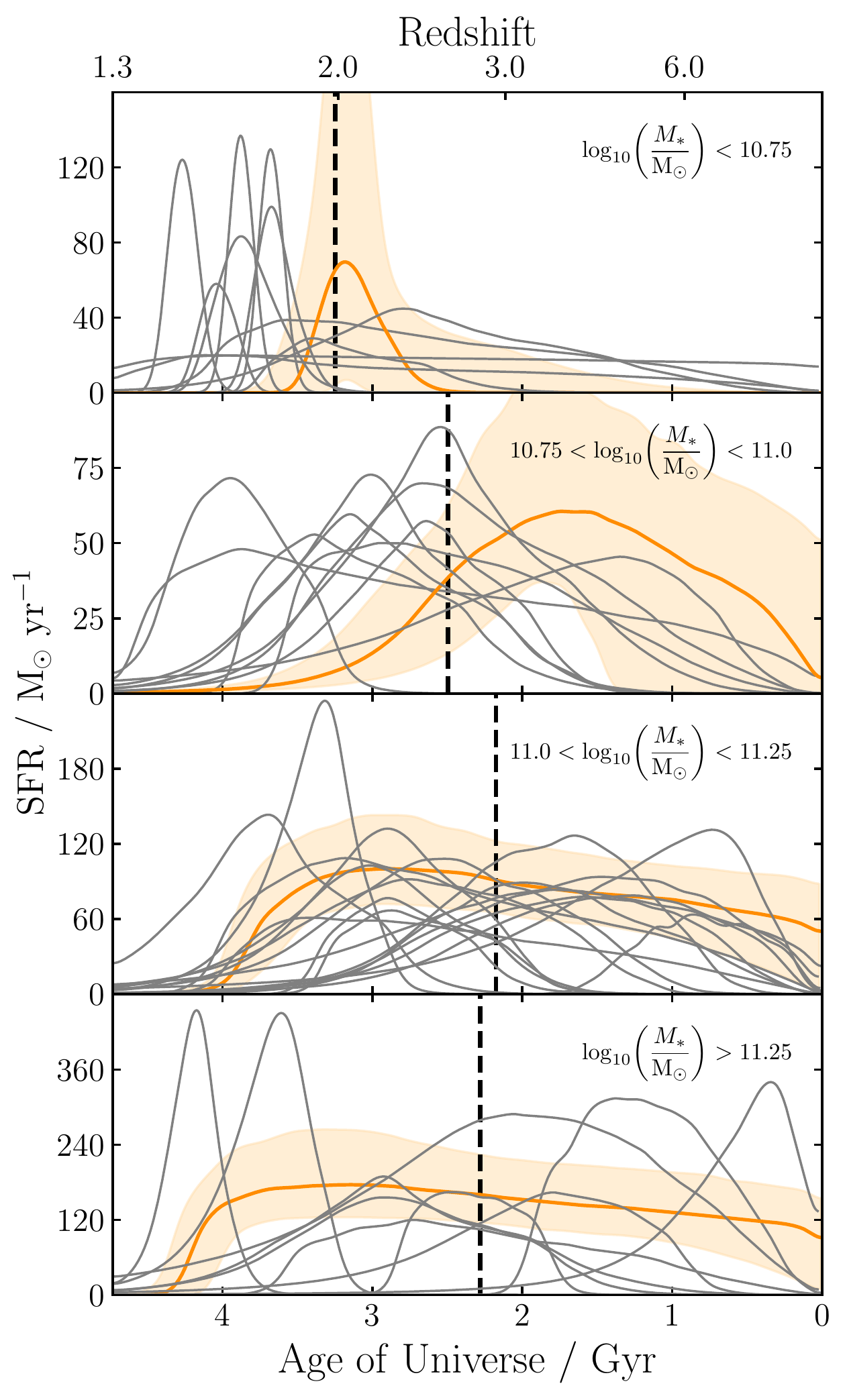}
    \caption{Posterior median SFHs for our quiescent sub-sample, divided into four bins in stellar mass. In each bin, a randomly selected SFH is highlighted, and the $16^\mathrm{th}{-}84^\mathrm{th}$ percentiles of the posterior are shaded to demonstrate the typical uncertainties. The dashed vertical lines show the mean formation times for galaxies in each bin, as described in Section \ref{subsect:results_age} and shown in Fig. \ref{fig:ages}.}\label{fig:sfh_post}
\end{figure}

We interpret these findings in the context of the results of \cite{Wild2016}, \cite{Almaini2017} and \cite{Maltby2018}, who suggest a dual origin for the post-starburst population. High-redshift PSBs primarily follow a UVJ evolution similar to that suggested by \cite{Belli2018}. They form in a brief, intense starburst, as suggested by \cite{Wild2016}, but either never build up significant dust or shed their dust rapidly as star-formation begins to fall. This could be a consequence of strong AGN-driven outflows at high-redshift (e.g. \citealt{Maiolino2012}; \citealt{Cimatti2013}), which have been linked to $z\sim1$ PSBs (\citealt{Maltby2019}). These galaxies enter the PSB box by the bottom-left (or possibly bottom-right) edge, and exit by the top-right. The morphological evidence presented by \cite{Almaini2017} and \cite{Maltby2018} supports this picture, with high-redshift PSBs found to be extremely compact, consistent with a major disruption event such as a merger.

By contrast, lower-redshift PSBs primarily form by the rapid quenching of normal star-forming galaxies \citep{Wild2016}, which shed their dust more slowly, following a UVJ evolution similar to that shown in the bottom-left panel of Fig. \ref{fig:uvj_tracks}. These objects briefly enter and leave the PSB box by the top-right edge. This is again consistent with \cite{Maltby2018}, who find that these low-redshift PSBs have less-concentrated structures, more similar to ordinary star-forming galaxies. At progressively lower redshifts, the progressively older stellar populations present in these objects prevent them from falling as far into the PSB box, explaining the decreasing number density of PSBs with redshift. This also explains the clustering of galaxies close to the top-right edge of the PSB box (e.g. see fig. 10 of \citealt{Belli2018}), which becomes more pronounced at lower redshifts.

A final piece of evidence for this scheme comes from the individual SFHs we infer for the three spectroscopic PSBs identified in Fig. \ref{fig:oii}. The posterior SFHs we infer for these objects are shown in Fig. \ref{fig:psb_sfhs}. For two objects the SFHs we infer are very extended before rapidly quenching. These objects are consistent with the cusp of the track shown in the bottom-left panel of Fig. \ref{fig:uvj_tracks}. Whilst rapidly quenched, these objects are not, in a literal sense, ``post-starburst". By contrast, the third object {\it is} a literal post-starburst, with almost all its stellar mass formed within a $\lesssim500$ Myr period, approximately 1 Gyr before it is observed. This object has the youngest mass-weighted age of the galaxies in our quiescent sub-sample. It can be seen in Fig. \ref{fig:uvj} to the extreme lower left of the quiescent population, and is the spectroscopic PSB closest to the \cite{Belli2018} selection box. It also has the strongest H$\delta$ absorption within our quiescent sub-sample (see Fig. \ref{fig:oii}). Further discussion of the SFHs and physical properties of VANDELS $z\sim1$ PSBs will be presented by Wild et al. in prep.

Our model galaxies which quench more slowly can be seen to follow a more conventional path in Fig. \ref{fig:uvj_tracks}, crossing the green valley approximately perpendicular to the edge of the UVJ selection box, and joining the red sequence higher up. These tracks are plausibly linked to  objects higher up the red sequence in Fig. \ref{fig:uvj} which still retain higher $A_\mathrm{V}$. However, most of their dust has already been lost by the time they cross the dashed line, preserving the strong age trend along the red sequence found by \cite{Belli2018}.

\subsection{The star-formation histories of massive quiescent galaxies at 1.0 < z < 1.3}\label{subsect:discussion_sfhs}

In light of the discussion of Section \ref{subsect:discussion_sequence}, it is interesting to consider the details of the SFH shapes we find for our quiescent sub-sample. The posterior median SFHs we infer are shown in Fig. \ref{fig:sfh_post}, split into the same four mass bins used in Section \ref{subsect:results_age}. To demonstrate the magnitude of the typical uncertainties, in each bin a randomly selected object is highlighted in orange, and the $16^\mathrm{th}{-}84^\mathrm{th}$ percentiles of the posterior are shaded. The mean formation times for galaxies in each bin (corresponding to the black errorbars on the top panel of Fig. \ref{fig:ages}) are shown with black dashed vertical lines.

The average trend towards earlier formation with increasing stellar mass shown in Fig. \ref{fig:ages} is also visible in Fig. \ref{fig:sfh_post}, however there is significant variation within each mass bin. In particular, in the highest mass bin, the SFH shapes are extremely diverse, which results in the average formation time being later than the next lowest mass bin. This demonstrates that, even at fixed stellar mass and observed redshift, the quiescent population contains galaxies with a wide range of formation histories.  The fact that knowledge of the stellar mass is not sufficient to make strong predictions about the SFH suggests again that a range of physical processes contribute to the quenching of star-formation.

In our lowest-mass bin the majority of objects formed recently and have very bursty SFHs with short $\lesssim 500$ Myr formation timescales. A smaller number of similar objects are also present in our higher-mass bins, though they are a minority. The PSB shown in the bottom panel of Fig. \ref{fig:psb_sfhs} is the youngest member of this group, though other members form later in cosmic time, as they have lower observed redshifts. Many or all of these objects are likely to have experienced a PSB phase at an earlier stage in their evolution. They are also plausibly linked with the population of highly star-forming submillimetre galaxies, the redshift distribution of which peaks at $z\sim2$ (e.g. \citealt{Dunlop2017}). The number of such objects present in our VANDELS DR2 sample does not allow us to place strong statistical constraints on their number density as a function of redshift. However the full VANDELS sample will contain approximately twice as many objects consistent with our selection criteria.

The older galaxies in our sample display a broad range of quenching timescales. Those which quench rapidly are likely to have experienced both green-valley and PSB phases, consistent with our ``early fast quench" model in Fig. \ref{fig:uvj_tracks}. A fraction however also have SFHs consistent with our ``early slow quench" model, with quenching timescales of $\gtrsim1$ Gyr. The two oldest galaxies in our sample, both of which have log$_{10}(M_*$/M$_\odot$) > 11, quenched at $z > 3$. This corresponds to a comoving number density of $2\pm1\times10^{-5}$ Mpc$^{-3}$ for quiescent galaxies at $z=3$, consistent with \citealt{Schreiber2018}, who find a number density of $1.4\pm0.3 \times 10^{-5}$ Mpc$^{-3}$ for quiescent galaxies with log$_{10}(M_*$/M$_\odot$) > 10.5 at $3 < z < 4$.

The ``early slow quench" objects may genuinely have experienced slow quenching at the highest redshifts, however two other scenarios are possible. Firstly, their SFHs may be composites of several systems which formed and quenched at different times, then merged to form the most massive quiescent galaxies. Secondly, it is possible that these systems were rapidly quenched, however, because their stellar populations are already old when observed, and hence slowly evolving, our observations cannot rule out slower quenching scenarios. Future instruments, such as NIRSpec on board the {\it James Webb Space Telescope} (JWST), which hold the promise of extending VANDELS-quality observations to quiescent galaxies at the highest redshifts, will provide invaluable constraints on the earlier evolution of these systems.

\section{Conclusion}\label{sect:conclusion}

In this work we report SFHs for a sample of 75 UVJ-selected galaxies with stellar masses of log$_{10}(M_*$/M$_\odot$) > 10.3 at observed redshifts of $1.0 < z < 1.3$. As described in Section \ref{sect:data}, our data consist of deep rest-frame UV spectroscopy from VANDELS, as well as multi-wavelength photometry. Using the \bagpipes\ code we fit our combined datasets with a sophisticated joint model for the physical properties of our galaxies and systematic uncertainties affecting our spectroscopic data, described in Sections \ref{sect:model} and \ref{sect:fitting}. The combination of extremely deep VANDELS spectroscopy with our sophisticated fitting methodology allows us to significantly improve upon previous analyses, obtaining strong, yet realistic constraints on the SFHs of our target galaxies.

We firstly quantify the average stellar mass vs stellar age relationship for massive quiescent galaxies at $1.0~<~z~<~1.3$ (Section \ref{subsect:results_age}). We find a steep trend towards earlier mass-weighted formation times with increasing stellar mass (downsizing) of $1.48^{+0.34}_{-0.39}$ Gyr per decade in mass (Equation \ref{eqn:tform}). We observe some evidence for the flattening of this trend at the highest masses ($M_* > 10^{11}\mathrm{M_\odot}$), as was reported by \cite{Gallazzi2014}. A slightly shallower trend of $1.24^{+0.27}_{-0.30}$ Gyr per decade in mass is observed for $r$-band light-weighted formation times (Equation \ref{eqn:trband}). As shown in Fig. \ref{fig:tform_spec}, the slope of this relationship is in agreement with spectroscopic results at $z\sim0.1$ and $z\sim0.7$ from \cite{Gallazzi2005, Gallazzi2014}, and at $z\sim1.75$ from \cite{Belli2018}. 

Recent photometric studies find weaker trends, of $\lesssim0.5$ Gyr per decade in mass (\citealt{Pacifici2016}; \citealt{Carnall2018}; see Fig. \ref{fig:tform_phot}), which we attribute to larger uncertainties in individual age determinations due to the age-metallicity-dust degeneracy (see Section \ref{subsubsect:discussion_downsizing_photometry}). We conclude that, at fixed observed redshift, an evolution in the stellar age vs stellar mass relationship of $\sim1.5$ Gyr per decade in mass is a robust result, which holds across the observed redshift interval from $0 < z < 2$.

This result places strong constraints on the AGN-feedback models used in modern cosmological simulations. As such, in Fig. \ref{fig:tform_spec}, we compare predictions from the \textsc{Simba} and \textsc{IllustrisTNG} simulations to our observational results. We conclude that the stellar mass vs stellar age relationships predicted by these simulations at $z=0.1$ are in good agreement with observations. However, at $z=1$, the relationships predicted are considerably flatter than our observational results, at $\lesssim0.5$ Gyr per decade in mass. This conclusion supports recent findings, which suggest that, whilst modern simulations now reproduce well the properties of local massive quiescent galaxies (e.g. \citealt{Dave2017}; \citealt{Nelson2018}), agreement is increasingly poor with increasing observed redshift (e.g. \citealt{Schreiber2018}).

By considering the distributions we find for galaxy physical parameters on the UVJ diagram (Fig. \ref{fig:uvj}), and the SFH shapes we infer for our sample (Fig. \ref{fig:sfh_post}), we attempt to understand the connection between green-valley, post-starburst (PSB) and quiescent galaxies, and to constrain quenching mechanisms at $z>1$. We demonstrate that typical green-valley galaxies, if rapidly quenched, pass through a PSB phase en route to quiescence, and show that SFHs consistent with this evolution exist within our sample. 

We additionally identify a class of predominantly $\mathrm{log}_{10}(M_*/\mathrm{M_\odot}) \sim 10.5$ galaxies which formed and quenched at $z<2$ in extreme starbursts with $\lesssim500$ Myr timescales. These objects are consistent with mergers and associated strong, AGN-driven outflows, and are plausibly related to submillimetre galaxies. These objects also pass through a PSB phase, supporting a dual origin for the PSB population (e.g. \citealt{Wild2016}; \citealt{Almaini2017}; \citealt{Maltby2018}). We finally find that some of our most massive galaxies appear to exhibit slow quenching at early times ($z>2$), though their quenching timescales are harder to constrain due to their older stellar populations. These objects are plausibly the result of mergers between galaxies with stellar populations formed at different times. To understand the earlier evolution of these systems in detail, deep continuum spectroscopy must be extended to the highest redshifts.

Our results demonstrate the power of large, high-redshift spectroscopic surveys for placing strong constraints on the evolution of subtle galaxy physical parameters across cosmic time. These results are important for furthering our understanding, as they are strongly constraining on models of galaxy formation. Upcoming instruments such as NIRSpec on JWST and MOONS at the VLT will greatly expand our high-redshift spectroscopic capabilities, and sophisticated fitting methodologies, such as presented in this work, will be key to realising their full potential.

\section*{Acknowledgements}

The authors would like to thank Philip Best, Dylan Nelson and Po-Feng Wu for helpful discussions which have improved the quality of this work. A. C. Carnall, F. Cullen and S. Appleby acknowledge the support of the UK Science and Technology Facilities Council. A. Cimatti acknowledges the grants PRIN MIUR 2015, ASI n.I/023/12/0 and ASI n.2018-23-HH.0. This work is based on data products from observations made with ESO Telescopes at La Silla Paranal Observatory under ESO programme ID 194.A-2003(E-Q). This work is based in part on observations made with the Spitzer Space Telescope, which is operated by the Jet Propulsion Laboratory, California Institute of Technology under a NASA contract. This research made use of Astropy, a community-developed core Python package for Astronomy \citep{Astropy2013}.

\bibliographystyle{mnras}
\bibliography{carnall2019} 

\bsp
\label{lastpage}
\end{document}